\documentclass[%
 reprint,
 superscriptaddress,
 showpacs,
 amsmath,amssymb,
 pre
]{revtex4-2}
\usepackage{setspace}
\usepackage{verbatim}
\usepackage{lmodern}
\usepackage{siunitx}
\usepackage{etoolbox}
\usepackage{booktabs}
\usepackage{tabularx}
\usepackage{makecell}
\usepackage{graphicx}
\usepackage{dcolumn}
\usepackage{bm}
\usepackage{bbm}
\usepackage{bbold}
\usepackage{physics}
\usepackage{enumerate}
\usepackage[dvipsnames]{xcolor}
\usepackage{times}
\definecolor{darkblue}{HTML}{023e8a}
\definecolor{darkred}{HTML}{780000}
\definecolor{darkgreen}{HTML}{005f73}
\definecolor{darkcyan}{HTML}{2a9d8f}
\usepackage[colorlinks=true,
	    urlcolor=darkred,
	    citecolor=darkblue,
	    linkcolor=darkred
            ]{hyperref}

\begin{document}

\title{Resonances of recurrence time of monitored quantum walks}
\thanks{\textcolor{darkcyan}{Invited contribution for the Special Topic: Festschrift for Abraham Nitzan, of the Journal of Chemical Physics. Dedicated to the 80-th Brithday of Abraham Nitzan}}

\author{Ruoyu Yin}
\email{yinruoy@biu.ac.il}
\author{Qingyuan Wang}
\email{qingwqy@gmail.com}
\thanks{equal contribution to this work.}
\affiliation{Department of Physics, 
Institute of Nanotechnology and Advanced Materials, 
Bar-Ilan University, Ramat-Gan 52900, Israel}

\author{Sabine Tornow}
\affiliation{Research Institute CODE, University of the Bundeswehr Munich, 81739 Munich, Germany} 
 
\author{Eli Barkai}
\affiliation{Department of Physics, 
Institute of Nanotechnology and Advanced Materials, 
Bar-Ilan University, Ramat-Gan 52900, Israel}

\begin{abstract}

The recurrence time is the time a process first returns to its initial state. Using quantum walks on a graph, the recurrence time is defined through stroboscopic monitoring of the arrival of the particle to a node of the system. When the time interval between repeated measurements is tuned in such a way that eigenvalues of the unitary become degenerate, the mean recurrence time exhibits resonances. These resonances imply faster mean recurrence times, which were recorded on quantum computers. 
The resonance broadening is captured by a {restart uncertainty relation}
[R. Yin, Q. Wang, S. Tornow, E. Barkai, Proc. Natl. Acad. Sci. U.S.A. 122, e2402912121 (2025)].
To ensure a comprehensive analysis, we extend our investigation to include
the impact of system size on the widened resonances, showing how the connectivity and energy spectrum structure of a system influence the restart uncertainty relation.
Breaking the symmetry of the system, for example time-reversal symmetry breaking with a magnetic flux applied to a ring, 
removes the degeneracy of {the eigenvalues of the unitary}, hence modifying {the mean recurrence time and the widening of the transitions}, and this effect is studied in detail.
The width of resonances studied here is related to the finite time resolution of relevant experiments on quantum computers, and to the restart paradigm.
%
\end{abstract}
\maketitle

\section{Introduction}

The concept of first hitting time, or first passage time, 
which represents the time it takes for a stochastic process to reach a specific target state for the first time, has been extensively explored across various fields,
such as physics, computer science, chemistry, biology, and finance \cite{Redner2001}.
In the realm of quantum physics, 
the investigation of quantum hitting times has attracted significant attention, 
offering valuable insights into fundamental aspects of measurement theory and quantum information
\cite{ALLCOCK1969,Kumar1985,Krovi2006,Portugal2010,Magniez2008,Gruenbaum2013,
Dhar2015a,Dhar2015,Friedman2017a,Tang2018exp,Felix2018,Lahiri2019,yin2019,Thiel2020D,Thiel2021entropy,David2021,Ziegler2021,dubey2021quantum,Didi2022,Liu2022a,Liu2023a,Yajing2023,Zhenbo2023,walter2023thermodynamic,roy2025causality}.
Quantum hitting times become particularly intriguing 
when examined in the context of mid-circuit measurements on quantum computers \cite{Sabine2022, Wang2023, yin2024restart}.
These measurements, taken at predetermined intervals with unitary dynamics between them, 
give rise to inherently random quantum trajectories.
Once these trajectories are established, 
we can study the random variable representing the time required to reach the desired target state. 
The behavior of quantum walk hitting times is vastly different if compared with hitting times generated with classical random walks, in particular the phenomenon of resonances studied here.

A profound {resemblance} exists between quantum hitting times and spectroscopy.
Consider, for instance, the interaction between a hydrogen atom and a periodic external field.
Under resonance conditions, where the periodicity of the field matches the energy difference of a transition,
we observe quantized transitions between energy levels \cite{cohen2019quantum1,cohen2019quantum2,hollas2004}.
{Somewhat similarly}, {for the recurrence problem (defined below)} the mean of quantum hitting times exhibits quantization \cite{Gruenbaum2013,Friedman2017a,David2021}.
In traditional spectroscopy, as in optics, transitions involve the creation or destruction of a photon,
particularly in weak external fields.
In contrast, in the quantum hitting time problem, no photons are involved;
instead, a dark state emerges upon transitions \cite{Thiel2020D}.
{Mathematically, the creation of dark states implies that a certain eigenvalue of the survival operator, defined below, approaches unity. This particular eigenvalue is crucial for our study and is discussed below in detail.}
Furthermore, {while the periodicity of the external field is driving resonances in many cases, here it is the periodicity of projective measurements, 
which in turn imply an overall non-unitary dynamics.
In our recent publication we examined the broadening effect of resonances of recurrence times, promoting a new uncertainty-like relation.}

{Remarkably, the observation of the spectroscopy-like resonance broadening
is closely related to the concept of restart, 
which has been extensively investigated both for classical 
\cite{Majumdar2011,Pal2016,restart,Igor2018,Majumdar2020,Gupta2022,
Ralf_Sandev2022,Ralf_WeiWang2022,barkai2023reset,Ralf2024competing,
Pal2023stocha,Pal2024Active,yael2023,linn2024cover,radice2024optimal} 
and quantum systems  \cite{Majumdar2018,Rose2018,Belan2020,riera2020measurement,Perfetto2021,Perfetto2021a,Turkeshi2021,das2022quantum,Magoni2022,Modak2023nonhermitian,gupta2023tight,Majumdar2023a,Majumdar2023b,chatterjee2023quest,Ruoyu2023,yin2023b,yin2024restart,Wald2024stochastic,Shukla2024accelerated}.
Restart is conventionally employed to accelerate search processes 
or to generate novel steady states \cite{Majumdar2011,Eule2016}.
Here, we emphasize the significance of restart in the context of hitting times and repeated measurements. 
Since any practical laboratory experiment does not allow for 
infinitely long measurement times, 
the approach of restart is naturally employed in the laboratory,
which addresses finite resolution effects.
Typically, finite resolution introduces only minor deviations from theoretical predictions.
However, in our context, the restart paradigm becomes particularly significant 
due to the discontinuous nature of the theoretical mean recurrence time. 
Even when the restart interval is extremely long, its effect cannot be disregarded.
}

{In our study \cite{yin2024restart}, we established a general formalism, 
a {restart uncertainty relation} that relates the finite duration of the experiment, 
the variance of the recurrence time, and the width of resonances. 
We also explored a time-energy relation for the problem, which is discussed below. 
However, key aspects of these general relations remain to be studied. 
Here, we address this gap, at least partially, 
by examining the effect of system size on the broadening. 
The fundamental question is whether increasing system size 
will reduce or increase the width of resonances, and under what conditions. 
We also investigate the effect of time-reversal symmetry breaking 
by considering a model of a quantum walk on a ring threaded by a magnetic field. 
This symmetry breaking removes degeneracies from the energy spectrum 
and can be used to control the resonances.}

The rest of the paper is organized as follows:
in Sec. \ref{sec2} we present the model,
and Sec. \ref{sec3} gives a recap of the quantum hitting problem. 
With the implication of restart approach demonstrated in Sec. \ref{sec4}, 
in the context of monitored quantum dynamics,
we uncover the {restart uncertainty relation} in Sec. \ref{sec5} and Sec. \ref{sec8},
which are briefly summarized in \cite{yin2024restart} with quantum computer implementation.
Specific examples are considered to validate our theory in Sec. \ref{sec7}.
{These examples include graph models that exceed the current capabilities 
of quantum computer simulations with repeated mid-circuit measurements, 
as well as cases involving a magnetic field. 
The latter demonstrates that our {restart uncertainty relation} remains valid irrespective of time-reversal symmetry.}
%
{To elucidate the impact of system energies-whose spectra 
are sensitive to system size-on resonance broadening, 
we examine the size dependence of our {restart uncertainty relation} in Sec. \ref{SizeEffects}.
Our analysis reveals non-universal behaviors that depend on the choice of resonances and the connectivity of graph models.}
We close the paper with conclusions in Sec. \ref{sec9}.
% ...





\section{Model}\label{sec2}

We consider a continuous-time quantum walk \cite{Farhi1998,Andrew2004,Solenov2006,Agliari2010,Blumen2011,WangKK2020,ximenes2023parrondos} 
governed by a tight-binding Hamiltonian for a finite graph, 
and the localized spatial states are $\{ \ket{x} \}$.
The system Hamiltonian is time-independent and denoted $H$, 
further we use $\hbar=1$, and hence as usual
the measurement-free evolution is defined via unitary dynamics 
$\hat{U}(\tau)= e^{- i H \tau}$.
We start with a recap on the hitting time problem, 
and later we introduce the restart. 

To define the quantum first hitting time, 
we employ the stroboscopic measurement protocol 
\cite{Krovi2006,Gruenbaum2013,Dhar2015,Dhar2015a,Friedman2017a,Felix2018,Thiel2020D,Thiel2021entropy,yin2019,David2021,Ziegler2021,Liu2022a,Liu2023a,Didi2022,Yajing2023,dubey2021quantum,Zhenbo2023,Lahiri2019,walter2023thermodynamic}.
Specifically, we initialize the system at the state $\ket{\psi_\text{in}}$,
which can be a node on the graph.
Every $\tau$ unit of time we perform a measurement. 
As for classical hitting times, 
it is natural to study the first time the system arrived at a target state.
The target state will be denoted $\ket{\psi_\text{d}}$, 
with the subscript $\text{d}$ for detected. 
Typically this will be a node on the graph. 
The measurements are based on the collapse postulate, 
in fact the measurements on the quantum computer test this assumption \cite{Wang2023, yin2024restart}.
Hence the repeated strong measurements are modelled by the projection operator $\hat{D}=|\psi_\text{d} \rangle \langle\psi_\text{d}|$. 
Those repeated measurements interrupt the unitary dynamics of the system, 
at times $\{\tau, 2\tau, 3\tau, \cdots \}$.  
$\tau$ is the sampling time or measurement period,  
a control parameter which can be modified in experiments. 
In some sense the repeated stroboscopic measurements performed on the system
emulate the role of a periodic force exerted on some material system, 
akin to the application of alternating fields in the realm of spectroscopy,
though now the effect is non-unitary \cite{Dhar2015a,Lahiri2019,Felix2020,dubey2021quantum}.

The output of the protocol is a string of measurement outputs of yes or no, concerning whether the quantum walker is detected or not.
Let $n$ be the first event when a click yes was recorded,
thus $n \tau$ is the first hitting time of the target state. 
Clearly, $n$ is random according to the nondeterministic nature of quantum mechanics.
For example, one may get in some experiment trial, \{no,no,no,yes\}, 
indicating the first hitting time is $4\tau$ or $4$ in units of $\tau$,
and in another attempt the first hitting time could be $10$ in units of $\tau$, etc.
Then one can obtain the mean time, $\expval{n}$, 
via performing an ensemble average of a large number of trials.

\begin{figure*}[ht]
\centering
\includegraphics[width=1.0\linewidth]{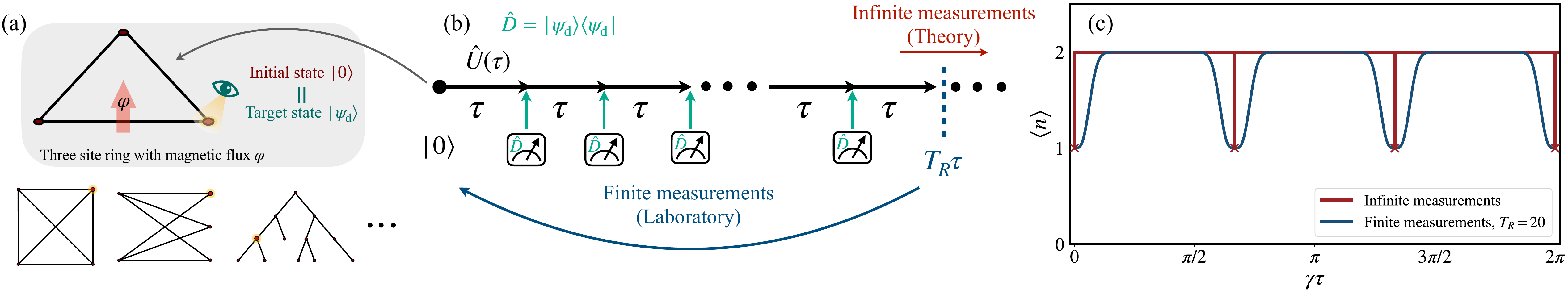}
\caption{{\bf Measurement protocol and broadening effect.}
{The theory presented in this paper is general and applicable to quantum walks on graphs. 
For demonstration purposes, 
panel (a) shows a three-site ring model with a magnetic flux $\varphi$,
which serves as our primary example throughout this figure.}
(b) Measurement protocol in theory and laboratory. 
We note in practice there exits a maximum measurement time $T_R$, beyond which one has to restart the process.
(c) The mean quantum first return time of the three-site ring model {without magnetic flux} in theory (red line) 
and in laboratory (blue line).
The red line shows the exact theoretical results for $T_R = \infty$ (from Eq. (\ref{meanw})),
exhibiting discontinuous transitions of $\expval{n}=w$ from $w=2$ to $w=1$ 
at $\gamma\tau=2\pi k /3$ ($k=0,1,2,\dots$).
The blue line displays numerical results for finite $T_R = 20$, 
demonstrating the broadening effect observed in laboratory measurements.
We analyze this broadening effect in detail and show its connection 
to an uncertainty-like relation.
For experimental implementations, see Refs.~\cite{Wang2023, yin2024restart}.
}
\label{fig:Theomean}
\end{figure*}

While the {restart uncertainty relation} presented in this work is generally valid,
in examples we will consider graph models governed by the 
Hamiltonians related to the adjacency matrix $A$ of the graph,
\begin{equation}
\label{eq0}
H = -\gamma A,
\end{equation}
where $\gamma$ is the hopping amplitude.
For our first example, a ring model will be utilized, see Eq. (\ref{eq1}) below. 
%
%
More specifically, we will consider 
a triangle and a benzene-like ring,
as paraphrastic models due to their simple energy spectra.
We will also investigate the effect of a magnetic field on the {restart uncertainty relation},
with an additional parameter, magnetic flux, tunable for the investigation of resonances in the system.
The target state will be a node on the graph. 
We will soon discuss the initial condition.
%
{Fig.~\ref{fig:Theomean}(a) illustrates several graph models to be examined in this work, 
with particular emphasis on the three-site ring with a magnetic flux. While we use this model as our primary example for theoretical explanation, 
our framework extends beyond graphs to encompass general quantum systems.}
%

%

\section{Recap: recurrence time problem}\label{sec3}
Grünbaum {\em et al.} \cite{Gruenbaum2013} studied the statistics of hitting times,
for the case when the initial state and the detected one are identical,
thus $\ket{\psi_{{\rm d}}}=\ket{\psi_{{\rm in}}}$.
For example if we start on a localized state on the graph, 
the first hitting time is the time the particle is first detected on its initial position, and hence in a sense the quantum walk is creating a loop, 
as the end state is the same as the initial one.
This problem is called the return or recurrence problem. 
Grünbaum {\em et al.} showed the relation of this problem to topology obtaining remarkable results.

Among other things they showed that with probability one 
the quantum walker is eventually detected. 
This holds for quantum walks, e.g. tight-binding quantum walks, 
in finite Hilbert space. 
Mathematically, let \(F_n\) be the probability of detecting the particle, 
at the target state, for the first time at the \(n\)-th measurement event, 
and specifically, $F_n$ can be obtained using \cite{Gruenbaum2013,Dhar2015a,Friedman2017a},
\begin{equation}
F_n = \left| \bra{\psi_\text{d}} \hat{U}(\tau){\cal S}^{n-1} \ket{\psi_\text{d}}\right|^2,
\label{Fn}
\end{equation}
where the survival operator ${\cal S} = ( 1-\hat{D} ) \hat{U}(\tau)$,
demonstrating the unitary evolution in the time interval $\tau$ 
followed by the complementary projection described by $ 1-\hat{D}$ 
(indicating null detection or the output no).
We then have \(\sum_{n=1}^{\infty}F_n = 1\). 
This statement, for a classical walker on a finite graph is trivial 
\cite{Polya1921}. 
In the quantum world it is not. 
If the initial state and the detected state are not identical, 
the eventual detection probability can be smaller than one even on a small graph \cite{Krovi2006,Thiel2020D}. 

Secondly, the authors in Ref. \cite{Gruenbaum2013} showed that 
the mean of the quantum return time is quantized 
\begin{equation}\label{meanw}
    \langle n \rangle = \sum_{n=1}^{\infty} n F_n = w.
\end{equation}
\(w\) is a winding number related to the generating function of the problem 
(see details in Appendix \ref{sec6}). 
It is a positive integer given by the number of distinct phase factors in the problem. 
More precisely, 
let \(\{ E_{1}, E_{2},\dots,E_k,\dots \} \) be the energy levels of the system, 
and \( \{ p_1, p_2, \dots ,p_k,\dots \} \) be the overlaps defined by 
\(p_k = | \langle E_k | \psi_\text{d} \rangle |^2\), 
where \(H\ket{E_k} = E_k\ket{E_k}\). 
Then for \(p_k \neq 0\), we count the number of the distinct phase factors 
\(\exp(- i E_k \tau)\) in the problem, and that gives \(w\). 
Hence, for a system with three non-degenerate energy levels, 
with \(p_k \neq 0 \), 
and when phase factors do not match, \(\langle n \rangle=3\). 
When a pair of phase factors match, we get $\expval{n}=2$, 
and if all three phase factors match, $\expval{n}=1$, 
hence modifying a parameter like $\tau$ 
we may witness transitions or resonances in the winding number $w$.


\section{The importance of restart}\label{sec4}
%
Notwithstanding the aforementioned theoretical results,
when we probe these issues {on a quantum computer \cite{Wang2023, yin2024restart}}, 
we find that the sharp discontinuous transitions are significantly broadened.
A demonstration of this effect is presented 
for a tight-binding model on a ring graph model;
see Fig. \ref{fig:Theomean}.
{The Hamiltonian reads
\begin{equation}
\label{eq1}
H = -\gamma \sum_{x=0}^{L-1} \left( \ket{x} \bra{x+1}+ \ket{x+1} \bra{x} \right),
\end{equation}
where 
$L$ is the size of the system. 
The periodic boundary condition is satisfied, i.e. $\ket{L}=\ket{0}$.}
The eigenvalues of the Hamiltonian are 
%
\begin{equation}
\label{eigenvalues1}
E_l= - 2 \gamma \cos\left( {2 \pi l \over L} \right),  \ l=0,1,2,\cdots,L-1.
\end{equation}
As an example, for $L=3$, there are $2$ distinct energy levels, $\{ -2\gamma, \gamma \}$, 
since one of the levels is doubly degenerate.
Hence, according to the theory and as seen in Fig. \ref{fig:Theomean}(c) (red line), 
for most $\gamma\tau$ besides pointwise discontinuous jumps, 
the mean $\expval{n}$ is equal to $2$, 
identical to the number of distinct energy levels 
(here one can show that overlaps $p_k$ are non-zero). 
However, there are some exceptional $\gamma\tau$'s at which $\expval{n}$ jumps to unity.
These exceptional $\gamma\tau$'s correspond to the cases 
when energy phase factors $e^{- iE_k\tau}$ match, 
i.e. $\gamma\tau=2\pi k/3$ with $k$ an integer.
Here we see $\expval{n} =1$. So in this system we have transitions from $w=2$ to $w=1$.
The case $\expval{n}=1$, corresponds to the situation 
where the wave function revives on the origin, 
and hence is detected with probability one in the first measurement.
The exact results for $T_R=\infty$ 
($T_R$ is the time span within which we collect data, see details soon),
namely the theory stated in last section,
are presented in Fig.~\ref{fig:Theomean}(c), 
as a straight red line $\expval{n}=2$ with dips.

However, in laboratory data is collected within a finite time span, 
say the measurement time is upper bounded by $T_R$ 
(see schematics in Fig. \ref{fig:Theomean}(b)),
the jump from $w=2$ to $w=1$ for the three-site ring system has a finite width (blue line in Fig. \ref{fig:Theomean}(c) as a simulation).
{IBM quantum computer experiments \cite{Wang2023, yin2024restart} appear similar to those presented in the blue line 
{(where a typical $T_R$ is $20$ when the experiment was conducted)}, 
and not to the theoretical result with infinite measurements (red line in Fig. \ref{fig:Theomean}(c)). 
We will show that this is due to finite time effects and the restart paradigm.}

%
To understand better the phenomenon, 
we now explain how the experiment is implemented in a laboratory
(see Ref. \cite{Wang2023} for a quantum computer realization). 
Clearly as in any other experiment to estimate the mean return time, 
we perform a finite number of measurements with an upper bound $T_R$. 
In other words, we have a maximum number of measurements $T_R$, 
beyond which we restart the process.
Obviously, in any experiment $T_R$ is finite.
Hence the measurements yield strings, 
for example, for \(T_R = 10\),
$$\{ \mbox{no},\mbox{no}, \mbox{yes}, \mbox{yes},\mbox{no}, \mbox{no}, \mbox{yes},\mbox{no}, \mbox{no}, \mbox{no}\},$$
no and yes stand for failure and success in detection, respectively. 
In principle, the experiment ends after the appearance of the first yes, 
so $n=3$ in this example,
namely the last \(7\) measurements in the above example are redundant, 
and only the first three bits are required. 


More importantly, since any experimental \(T_R\) is finite, 
we get a broadening of the topological transition found when $\langle n \rangle$ is estimated; see the blue line in Fig.~\ref{fig:Theomean}(c). 
Simply enough, non-analytical transitions, 
which are valid in principle when \(T_R \to \infty\), 
are not a possibility in experiments. 
Hence the restart problem and finite \(T_R\) resolution become interesting, 
when parameters are tuned close to the topological transition. 
This, as we will show, is related to an extremely slowly-relaxing mode. 
Since the discontinuous transitions of $\expval{n}$ are a quantum feature, related to the merging of phase factors $\exp(- iE_k\tau)$, the restart features found below, have no simple classical analogue.

The finite \(T_R\) implies that we may find a sequence 
with no detection at all in laboratory,
\begin{equation}\label{eqNO}
\{ \mbox{no}, \mbox{no}, \cdots , \mbox{no}\}.
\end{equation}
As mentioned after each sequence of length \(T_R\) we restart the process. 
{Sequences like Eq. (\ref{eqNO}) are rare, 
but close to the topological transition, 
the effect is crucial. 
This is because of the non-analytical nature of $\expval{n}$ 
in ideal systems when $T_R \to \infty$.}

More generally, the restart strategy was widely studied,
in the context of stochastic processes, 
for example in biochemical reactions, home-range searches of animals, etc.
\cite{pnas,Denis2014}. 
For example, a search expedition after a lost person,
may return to base at the end of the day, 
just to restart in the following days. 
The restart literature in the context of hitting times can be roughly divided into two. 
The first tries to optimize the first passage/hitting time, 
and show that restart is a useful strategy \cite{Redner2020,Ruoyu2023,chatterjee2023quest,biswas2024drift,Blumer2024a,Blumer2024b}. 
The second aims to study the effects of restart \cite{Eule2016,Perfetto2021,barkai2023reset,yin2023b},
for example due to limited resources on home-range searches, 
on the first hitting time. 
Our study is closer to the second line of thought. 
The novelty comes from the discontinuity of the recurrence times, 
which is a quantum feature.

\section{Restart uncertainty relation for the return time}\label{sec5}
{
In this section, following Ref. \cite{yin2024restart}, 
we derive the restart uncertainty relation and provide additional details 
on the variance of the recurrence time. 
The results here further elucidate the relationship among the restart time of measurement,
the uncertainty in the measured mean recurrence time 
and the fluctuations of the recurrence time.
}

The measured \(\langle n \rangle\) in laboratory,
presented in Fig.~\ref{fig:Theomean}(c), is  
\begin{equation}\label{cm}
    \langle n \rangle_{\text{Con}} = \frac{\sum_{n=1}^{T_R} nF_n}{\sum_{n=1}^{T_R} F_n}. 
\end{equation}
This is the mean of \(n\), 
conditioned that we detected the particle up to the \(T_R\)-th attempt.
The denominator indicates that the null-detection realizations, as shown in Eq. (\ref{eqNO}), are discarded.
Naively, in experiments, if the restart time (an integer) $T_R$  
is much larger than the theoretical mean $\langle n \rangle$ 
one would anticipate that $\expval{n}_{\text{Con}}$ will converge to $\langle n \rangle$  which is simply $w$.  
This is false close to the transition
as we have witnessed in Fig.~\ref{fig:Theomean}(c).
Note that we will soon study the restarted mean in Sec.~\ref{sec8}.

\begin{figure*}[t]
\begin{center}
\includegraphics[width=0.95\linewidth]{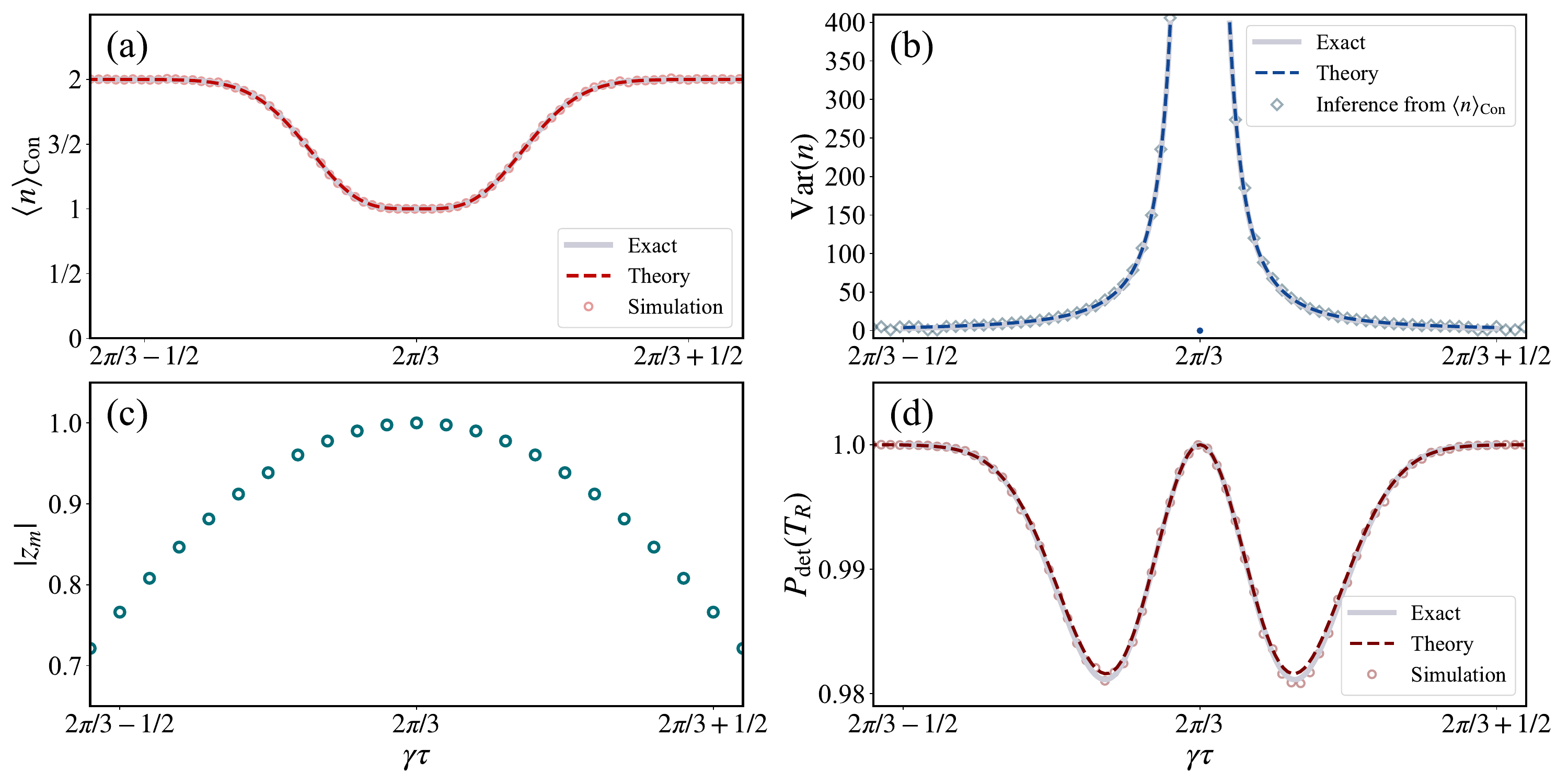}
\end{center}
\caption{
    (a) The transition from $\expval{n}_{\text{Con}} = 2$ to $\expval{n}_{\text{Con}} = 1$ and back is widened due to restarts. In particular here we restart after $T_R=20$ measurements. We compare the exact results obtained using Eqs. (\ref{Fn},\ref{cm}) with the theory Eq. (\ref{nvar}). 
    {The red circles represent the Monte Carlo simulation results, using the same parameters as in the theory and $300,000$ samples. These results align closely with the theoretical and exact values.}
    As shown in the text, as $T_R$ is increased the width will be diminished. 
    (b)  The variance of the quantum first hitting time 
    as a function of the control parameter $\gamma\tau$. 
    The variance diverges around the transition point $\gamma\tau = 2 \pi/3$. 
    The blue squares indicate the inferred variance from the Monte Carlo simulation results of $\expval{n}_{\text{Con}}$ using Eq. (\ref{nvar}). 
    The inferred results agree well with the theoretical and exact results.
    (c) The largest eigenvalue $|z_m|$ 
    of the survival operator ${\cal S}$ as a function of $\gamma\tau$. 
    {It is at its maximal value at the resonance.}
    {(d) The probability of detection $P_\text{det}(T_R)$ as a function of $\gamma\tau$. 
    The red circles represent the Monte Carlo simulation results with $T_R = 20$ and $300,000$ samples. The model used here is the same as in Fig. \ref{fig:Theomean}.}
}
\label{fig:Tridip}
\end{figure*}

We now consider in generality the \(w \to w-1\) transition and calculate the broadening of the resonances. 
We will show the phenomenon is related to an extremely slow-decay mode in the system. 
We assume that 
\begin{equation}
    F_n \sim a(z_m)|z_m|^{2n}, \;\; \text{when}\;n \to \infty, 
    \label{defFn}
\end{equation}
namely, the first detection probability decays exponentially \cite{Friedman2017a}.
Technically, $z_m$ is the largest zero of the generating function, 
or the complex conjugate of the largest eigenvalue of the survival operator ${\cal S}$ inside the unit circle, in the sense of its absolute value
\cite{Thiel2020D,Liu2022a}.
{Namely,
\begin{equation}\label{zmmax}
|z_m|=\max\left\{ |z_1|, |z_2|, |z_3|, \dots \right\},
\end{equation}
where \(m\) stands for maximum, 
and \(z_i\)'s are the complex conjugate of the eigenvalues of the survival operator ${\cal S}$.}
%
Following Refs. \cite{Gruenbaum2013,yin2019}, 
we know that in the vicinity of the transition, $|z_m|\to 1$, see Fig. \ref{fig:Tridip}, indicating slow relaxation. 
More precisely, at the transition itself, $|z_m|=1$ and then $\expval{n} = w-1$, 
while otherwise $|z_m|<1$ and $\expval{n} = w$, at least theoretically.
\(a(z_m)\) is to be determined, 
and luckily this is possible using simple hand-waving considerations 
related to the quantization of \( \expval{n} \). 
Recall that $|z_m|$ is controlled by the parameter $\tau$, 
as well as other details in the system.
Our goal is to find a relation between
$\expval{n}_{\text{Con}}$ and $z_m$.

Using the normalization \(\sum_{n=1}^{\infty} F_n = 1\), 
we consider the finite summation, i.e. the probability of detection within $T_R$ measurements
\begin{equation}
    P_\text{det}(T_R) := \sum_{n=1}^{T_R} F_n \simeq 1-a(z_m)\sum_{n=T_R+1}^{\infty}|z_m|^{2n},
    \label{sumfn}
\end{equation}
when \(T_R\) is large. 
Roughly speaking, \(T_R\) is large such that other modes are negligible. 
Note that here we are avoiding rigorous mathematics for simplicity,
see Appendix \ref{sec6} for a proof. 
Using a geometric series we have 
\begin{equation}\label{sumfna}
    P_\text{det}(T_R)
    \simeq 1- a(z_m)\frac{|z_m|^{2+2T_R}}{1-|z_m|^2}.
\end{equation}
%
The second term must be zero when \(T_R \to \infty\) 
and this holds true even if \(|z_m| \to 1\), 
so as we show later \(\lim_{|z_m| \to 1} a(z_m)/\left( 1-|z_m|^2 \right) = 0\). 
This means that as \(|z_m| \to 1\), 
the exponential decay of \(F_n\) is slower, 
however the prefactor \(a(z_m)\) is also becoming smaller, 
which makes sense since the \(F_n\)'s are normalized. 

To determine \(a(z_m)\) we exploit the quantization of \(\langle n \rangle\). 
Consider the summation 
\begin{equation}
    \sum_{n=1}^{T_R} n F_n = w - \sum_{n=T_R+1}^{\infty} n F_n,
\end{equation}
where we used 
\(\sum_{n=1}^{\infty} n F_n = w\), so \(|z_m| \neq 1\). 
Using Eq.~(\ref{defFn}), 
\begin{equation}
    \sum_{n=1}^{T_R} n F_n 
    \simeq w - a(z_m)|z_m|^{2+2T_R}
    \frac{1+T_R-T_R|z_m|^2}{\left( 1-|z_m|^2 \right)^2}. 
    \label{meanfin}
\end{equation}
Then if \(|z_m| = 1\) we have \(\sum_{n=1}^{T_R} nF_n \simeq w-1\), 
namely at \(|z_m|=1\), we have a jump of \(\langle n \rangle\) from \(w \to w-1\). 
We therefore argue that the second term in the right-hand side of Eq.~(\ref{meanfin}) 
is unity when \(|z_m| = 1\). 
Hence clearly we find in the limit \(|z_m|^2 \to 1\), 
\begin{equation}\label{azm}
    a(z_m) = \left( 1-|z_m|^2\right)^2. 
\end{equation}
With this formula we can find the effects of restart on the mean quantum hitting time. 

Since $|z_m|\to1$ in the vicinity of the transition, let 
\begin{equation}\label{eqeps}
|z_m|^2 = 1-\epsilon^2
\end{equation}
with \(\epsilon \to 0\). 
First, we find that Eq. (\ref{sumfn}) becomes
\begin{equation}\label{pdet15}
    P_\text{det}(T_R) =\sum_{n=1}^{T_R}F_n \simeq 1-\epsilon^2 e^{-T_R \epsilon^2},
\end{equation}
where the limit \(T_R \to \infty\) is considered 
and \( T_R \epsilon^2\) is finite.
Eq. (\ref{pdet15}) suggests that the probability of detection within $T_R$ measurements increases 
when the system is closer to the resonance, and reaches $1$ at resonance.
We will see this clearer later when considering specific systems.
Then in the same limit (\(\epsilon \to 0\), \(T_R \to \infty\)), 
with Eqs. (\ref{meanfin}-\ref{pdet15}) and Eq. (\ref{cm}), 
we obtain 
\begin{equation}
\begin{split}
        \langle n \rangle_{\text{Con}} 
        &=  \frac{ w - ( 1 + T_R \epsilon^2 ) e^{ -T_R \epsilon^2 }}
            {1-\epsilon^2 e^{-T_R \epsilon^2 }}.
\end{split}
\end{equation}
%
Thus, let us consider the limit that 
\begin{equation}\label{sigma}
    \sigma = T_R \epsilon^2
\end{equation}
is finite and find our first main result\cite{yin2024restart}
%
\begin{equation}\label{ntr}
    \boxed{
    \langle n \rangle_{\text{Con}} = 
    w - (\sigma + 1) e^{-\sigma}
    }. 
\end{equation}
This equation shows that if \(\sigma \to 0\) 
(namely we are on resonance, $|z_m|=1$, $\epsilon=0$), 
we have \(\langle n \rangle \to w-1\). 
If \(\sigma\) is large (namely \(|z_m|<1\) and finite while \(T_R \to \infty\)),
\(\langle n \rangle = w\), see Fig. \ref{fig:Tridip}(a) for \(w=2\) as an example. 
Hence this simple equation describes the transition between \(w\) and \(w-1\). 
However, so far we have only proposed some plausible math, 
the goal now is to connect \(|z_m|\) and hence \(\epsilon^2\) 
to physical parameters of the system, namely to fluctuations and to its energies. 

\subsection{$\langle n \rangle_\text{Con}$ is related to the fluctuations}

The fluctuations of the quantum return time are found by Grünbaum {\em et al.}, 
at least formally \cite{Gruenbaum2013}.
In the context of the first hitting time problem, 
the time for first detection is a random variable, 
and hence we have an operational meaning for the hitting time fluctuation, or the variance of $n$. 
Let the variance be 
\begin{equation}
    \text{Var}(n) = \langle n^2 \rangle - \langle n \rangle^2 
    = \sum_{n=1}^{\infty} n^2 F_n - \langle n \rangle^2,
\end{equation}
then 
\begin{equation}\label{varzeros}
    \text{Var}(n) = \sum_{i,j=1}^{w} \frac{2 z_i z_j^*}{1 - z_i z_j^*}.
\end{equation}
Here \(z_i\)'s are the zeros of the generating function, 
or the complex conjugate of the eigenvalues of the survival operator ${\cal S}$ inside the unit circle,
that are important for our discussion.
The computation of these \(z_i\)'s in general demands 
solving a polynomial equation of the order of the dimension of the Hilbert space, 
which will be discussed in the Appendix \ref{sec6} (see Eq.~(\ref{phiND})).
For now, the important issue is that \(0< |z_i| \leq 1\). 

Remarkably, in the vicinity of topological transitions, 
for example when \(w \to w-1\), 
i.e. when \(\langle n \rangle\) has a dip as a control parameter like $\tau$ is changed, 
the \(\text{Var}(n)\) diverges. 
These large fluctuation were studied in Ref. \cite{yin2019}. 
Under certain conditions one of the eigenvalues is approaching the unit circle, 
this being the largest eigenvalue in the system. 
We denoted this eigenvalue \(z_m\) as mentioned.
%
%
As stated in Eq.~(\ref{defFn}), \(z_m\) will control the large \(n\) limit of \(F_n\). 
Since \(|z_m| \to 1\), with Eq. (\ref{varzeros}), we have
\begin{equation}
    \text{Var}(n) \sim  \frac{2}{1-|z_m|^2}, 
    \label{eqvar}
\end{equation}
and hence, the fluctuations are large when \(|z_m|^2 \to 1\), see Fig. \ref{fig:Tridip}(b)(c).

Before discussing \(\langle n \rangle_{\text{Con}}\), 
let us first review $P_\text{det}(T_R)$ 
that characterizes the detection efficiency within the finite time span $T_R$, 
by efficiency we mean the proportion of total output strings 
which can be used to obtain the conditional mean hitting time.
Recall that $P_\text{det}(T_R) \simeq 1-\epsilon^2 e^{-\sigma}$ (Eq. (\ref{pdet15})),
and then with Eq. (\ref{eqvar}) we obtain
\begin{equation}
   \lim_{\epsilon\to0,T_R\to\infty} \text{Var}(n) \; [1- P_\text{det}(T_R)]
   = 2 \exp(- \sigma),
\end{equation}
%
where $\sigma = \epsilon^2 T_R$ is finite. 
Using Eqs.~(\ref{eqeps},\ref{eqvar}), we have 
\begin{equation}
    \text{Var}(n) \sim \frac{2}{\epsilon^2},
    \label{varE}
\end{equation}
and hence $\sigma = 2 T_R/\text{Var}(n)$. 
As mentioned before, close to the resonance $|z_m|=1$, 
the probability of {\em not} detecting until time $T_R$, $1- P_\text{det}$, 
goes to zero, and $\text{Var}(n)$ goes to infinity,
however, the multiplication of the two gives a nontrivial yet simple limit.

Now let us relate \(\langle n \rangle_{\text{Con}}\) 
with the fluctuations of the quantum first hitting time and the restart time $T_R$. 
%
Using $\sigma = 2 T_R/\text{Var}(n)$, Eq. (\ref{ntr}) becomes \cite{yin2024restart}
\begin{equation}\label{nvar}
    \boxed{
    \langle n \rangle_{\text{Con}} 
    =   w - \left[ \frac{2T_R}{\text{Var}(n)}+1 \right] 
        \exp \left[ -\frac{2T_R}{\text{Var}(n)} \right]
    }. 
\end{equation}
This is a relation between the mean and the uncertainty of fluctuations namely the variance. 
Unlike other uncertainty relations \cite{heisenberg1927,Robertson1929,Schroedinger1930,yin2019,Thiel2021entropy}, 
it is an equality (in the spirit of the energy-time relation, as shown below).
As mentioned, at the transition points \(\text{Var}(n) \to \infty\) (Fig. \ref{fig:Tridip}(b)), 
hence the measurement of \(\langle n \rangle_{\text{Con}}\) (Fig. \ref{fig:Tridip}(a)) 
provides direct information of \(\text{Var}(n)\). 
It is remarkable that a finite though large \(T_R\) measurement 
of the mean \(\langle n \rangle_{\text{Con}}\) 
yields information on the fluctuations, i.e. the \(\text{Var}(n)\) Eq.~(\ref{eqvar}),
where the latter are valid in the asymptotic limit.

The width of the transition is found for \(\langle n \rangle_\text{Con} = w-{1}/{2}\). 
Roughly speaking, this point marks an uncertainty in the value of the measured winding number.
Using Eq. (\ref{ntr}), we have
\begin{equation}\label{nsigma}
    (\sigma+1)\text{exp}(-\sigma) = {1 \over 2},
\end{equation}
%
%
%
which gives $\sigma \simeq 1.6784$.
Hence 
\begin{equation}\label{nvar167}
\boxed{
\begin{aligned} 
    \expval{n}_\text{Con} &= w-1/2, \\
    &\text{when }\, 2T_R /\text{Var}(n) = T_R (1-|z_m|^2) \simeq 1.6784
\end{aligned}}.
\end{equation}
This determines the restart time of measurement $T_R^*$ 
when the system is found at the midpoint of the transition,
\begin{equation}\label{wide21}
    T_R^* \simeq {1.6784 \over 2} { (\Delta t)^2 \over \tau^2},
\end{equation}
where $\Delta t = \tau\sqrt{\text{Var}(n)}$ is the standard deviation 
of the first hitting time.


\subsection{$\expval{n}_\text{Con}$ in terms of energies}
With Eq. (\ref{nvar}) 
and the results in Ref. \cite{yin2019}, we can re-express the conditional mean in terms of system parameters, say the energies.
As mentioned earlier, the topological transition is found when two phase factors merge. 
These correspond to two specific energies in the system, 
which we denote $E_+$ and $E_-$, namely $\exp(- i E_+ \tau) \simeq \exp(- i E_-  \tau)$ 
close to the transition, and other energies or phase factors are not playing a major role. 
Using the angular distance between the two phases, 
i.e. $\tau | E_+ - E_- | \mod 2 \pi =:\widetilde{\Delta E \tau }$, 
as a small parameter, and the perturbation method applied to the generating function, 
one can asymptotically express $z_m$, and then with Eq. (\ref{eqvar}), 
the variance of $n$ is quantified as \cite{yin2019}.
%
\begin{equation}\label{var23}
\begin{aligned}
    \text{Var}(n) 
    &\sim  2{ (p_+ + p_-)^3 \over p_+p_-} 
    {1 
    \over 
    \left( 
    \tau \left| E_+ - E_- \right| \,\,\, \text{mod}\,\, 2\pi
    \right)^2
    } 
    ,
\end{aligned}
\end{equation}
%
%
%
where the overlaps $p_\pm = \sum_l^{g_\pm} \abs{\braket{\psi_\text{d}}{E_{\pm,l}}}^2$, 
with $g_\pm$ the degeneracy of $E_\pm$. 
Eq~(\ref{var23}) is rewritten as
\begin{equation}\label{varlam}
    \text{Var}(n) \sim {2 \over \lambda} \left( \widetilde{\Delta E \tau } \right)^{-2},
\end{equation}
where $\lambda =  {p_+p_- / (p_+ + p_-)^3}$. 
Using Eq.~(\ref{varlam}), Eq. (\ref{nvar}) becomes\cite{yin2024restart}
\begin{equation}\label{condmean}
    \boxed{
    \expval{n}_\text{Con} 
    = w - 
    \left[ 1+ \lambda T_R (\widetilde{\Delta E \tau})^2 \right]
    \exp 
    \left[ -\lambda T_R (\widetilde{\Delta E \tau })^2 \right] }.
\end{equation}
%
%
Clearly, from Eq.~(\ref{condmean}), 
if $\widetilde{\Delta E \tau}=0$, 
i.e. $E_+ = E_-$ or $\tau | E_+ - E_- | = 2\pi k$ with $k$ an integer, 
we have $\expval{n}_{\text{Con}} = w-1$, otherwise and if $T_R\to \infty$, $\expval{n}_{\text{Con}} = w$. 
Thus the above equation described the transitions in terms of system parameters. 
Remarkably, it is only a few parameters, namely, all the energy phase factors 
besides the two that are approximately equal, are negligible.
In terms of the system parameters, 
we can also re-express Eq. (\ref{nvar167}) as
\begin{equation}\label{condhalf}
\begin{aligned}
    \expval{n}_\text{Con} &= w-1/2,  \\
    &\text{when }\, \lambda T_R 
    (\widetilde{\Delta E \tau})^2 \simeq 1.6784.
\end{aligned}
\end{equation}
This indicates more clearly the restart time $T_R$ at the half of the transition, inversely proportional to the square of the deviation from the resonance $\widetilde{\Delta E \tau}$.

In Fig. \ref{fig:Tridip} 
we demonstrate the approach using finite-measurement simulation.
Using a three-site ring model, 
we study the transition $\langle n \rangle =2\to \langle n \rangle= 1$. 
We show how Eq. (\ref{ntr}) 
agrees well with the numerics.
%
Hence in this case, i.e. the ring model with $L=3$, the width of the transition is characterized using 
\begin{equation}\label{eq33tri}
\begin{split}
    \expval{n}_\text{Con} &= 3/2, \\
    &\text{when }\,  2T_R (3\gamma\tau-2\pi k)^2/9 \simeq 1.6784 ,
\end{split}    
\end{equation}
%
with $k$ an integer.
One can reverse the thinking. 
Using Eq. (\ref{nvar}), 
one could infer a very precise value of $\text{Var}(n)$, 
for each $\gamma\tau$,  
from the data, given that $T_R=20$ in our numerical ``experiment''.
{In Fig.~\ref{fig:Tridip}(b), 
we also plot the variance $\text{Var}(n)$ (blue squares) 
directly inferred from the numerical data 
obtained with Monte Carlo simulations, 
which is in good agreement with the theory and exact results.
We also plot the maximum eigenvalue of the survival operator ${\cal S}$, versus $\gamma \tau$,  
in Fig.~\ref{fig:Tridip}(c). 
When this eigenvalue approaches unity we are at resonance, 
namely the dip of $\langle n \rangle_{\text{Con}}$ 
in Fig.~\ref{fig:Tridip}(a), signals the appearance of a dark state. Specifically, when $|z_m| = 1$, we create a dark state in Hilbert space.}


\section{Uncertainty-like relation for the restarted hitting time}\label{sec8}
As noted above using conditional measurements, 
we exclude sequences without any detection, 
see Eqs. (\ref{eqNO}) and (\ref{cm}).
Here we study the unconditional hitting time under restart, 
which is widely used in many applications \cite{Luby1993,pnas}.
Let $1\le  n_R  < \infty$ be the first hitting time in units of $\tau$, under restart, 
where the subscript $R$ is for restart. 
As usual $n_R$ is random. 
For example, consider the sequence of $\{\text{no},\dots,\text{no} \}$ of length $20$, 
which after restart, is followed by $\{\text{no}, \text{no}, \text{yes} \}$.
Here the first time for detection is
$23 \tau$ and $n_R = 23$. 
This restarted mean hitting time is given by \cite{Luby1993,Pal2021,Eliazar2021}
%
%
\begin{equation}\label{resmean}
    \expval{n_R} = \expval{n}_\text{Con} + {1-P_\text{det}(T_R)\over P_\text{det}(T_R)} T_R.
\end{equation}
%
We note that here we are using the so-called sharp restart at fixed non-random time $T_R$ \cite{Pal2016}, 
which was shown to be the most efficient way to generate the global minimum 
of mean hitting time under restart \cite{Luby1993,restart}.  


To deal with Eq. (\ref{resmean}) close to the transition $w\to w-1$, 
we again employ the asymptotic behaviors of $F_n$ Eq. (\ref{defFn}), 
and assume that $\epsilon^2\to0$, $T_R\to\infty$, and $\sigma = T_R \epsilon^2$ is finite.
Then with Eq. (\ref{pdet15}) and Eq. (\ref{ntr}), 
we can re-express Eq. (\ref{resmean}) in terms of $\sigma$,
\begin{equation}\label{resasymp}
\boxed{
    \expval{n_R} = w - e^{-\sigma}\,\,\,
    }.
\end{equation}
As expected, when $\sigma\to0$, $\expval{n_R}= w-1$,
and when $\sigma\to\infty$, $\expval{n_R}= w$.
Using $\sigma = 2 T_R/\text{Var}(n)$, 
this equation can be rewritten in terms of the fluctuations of quantum hitting times 
\begin{equation}\label{nrvar}
\boxed{
    \expval{n_R} = w - \exp \left[-2 T_R/\text{Var}(n) \right]
    },
\end{equation}
%
or in terms of energies
%
\begin{equation}\label{rmen}
    \boxed{
    \expval{n_R} 
    = w - 
    \exp 
    \left[ -\lambda T_R (\widetilde{\Delta E \tau })^2 \right] },
\end{equation}
where $\lambda$, $\widetilde{\Delta E \tau }$ have been defined before.
{Eqs. (\ref{resasymp}-\ref{rmen}) are remarkably simple, 
and they describe the topological transition. 
Namely, as we vary $\tau$,   
$\expval{n_R} = w$ far from resonance, 
and it is $w-1$ at resonances. 
The restart $\expval{n_R}$ is larger than $\expval{n}_\text{Con}$ as expected,
though at resonance and far from resonance they are identical.}

{As indicated by Eq. (\ref{resmean}), 
$P_\text{det}(T_R)$ is an important quantifier of the difference between the conditional mean $\expval{n}_\text{Con}$ 
and the restarted mean $\expval{n_R}$. 
Namely, using Eq. (\ref{pdet15}),
$\expval{n_R}-\expval{n}_\text{Con} = T_R [1-P_\text{det}(T_R)]/P_\text{det}(T_R) 
\sim \epsilon^2 T_R e^{-T_R \epsilon^2}$.
This suggests that the gap between $P_\text{det}(T_R)$ and $1$, 
amplified by $T_R$,
determines the difference between $\expval{n_R}$ and $\expval{n}_\text{Con}$.}
{We now specify the property of $P_\text{det}(T_R)$ 
around the resonance, to provide further insights.
For the three-site ring, with Eq. (\ref{eq33tri}), 
the $\epsilon^2$ in Eq.~(\ref{pdet15}) is 
\begin{equation}
\epsilon^2 = \lambda (\widetilde{\Delta E \tau})^2 = {2\over 9} (3\gamma\tau - 2\pi k)^2,
\end{equation}
with $k$ an integer. 
Hence, Eq.~(\ref{pdet15}) for $L = 3$ can be rewritten as
\begin{equation}
   P_\text{det}(T_R) \sim 
   1- {2\over 9} (3\gamma\tau - 2\pi k)^2 
   \exp \left[ - {2\over 9} (3\gamma\tau - 2\pi k)^2 T_R \right].
\end{equation}
We can see that, $P_\text{det}(T_R)\to 1$
when $3\gamma\tau - 2\pi k$ goes to $0$, 
i.e. when system parameters are approaching the resonance.
When we are far from resonance, 
namely $\widetilde{\Delta E \tau} \gg  1$, 
$P_\text{det}(T_R)$ is nearly unity, 
like the case $\widetilde{\Delta E \tau}=0$. 
Hence $P_\text{det}(T_R)$ versus $\tau$ will exhibit a $W$-like structure, as shown in Fig.~\ref{fig:Tridip}(d) where $T_R=20$ is used.
The red circles represent Monte Carlo simulation results 
with $300,000$ samples, 
the dashed line shows the theory
and the solid line indicates exact numerical results. 
All three approaches demonstrate nice agreement. 
Further discussion on $P_\text{det}(T_R)$ for other graph models 
will be provided later.}

\begin{figure*}
    \begin{center}
    \includegraphics[width=0.95\linewidth]{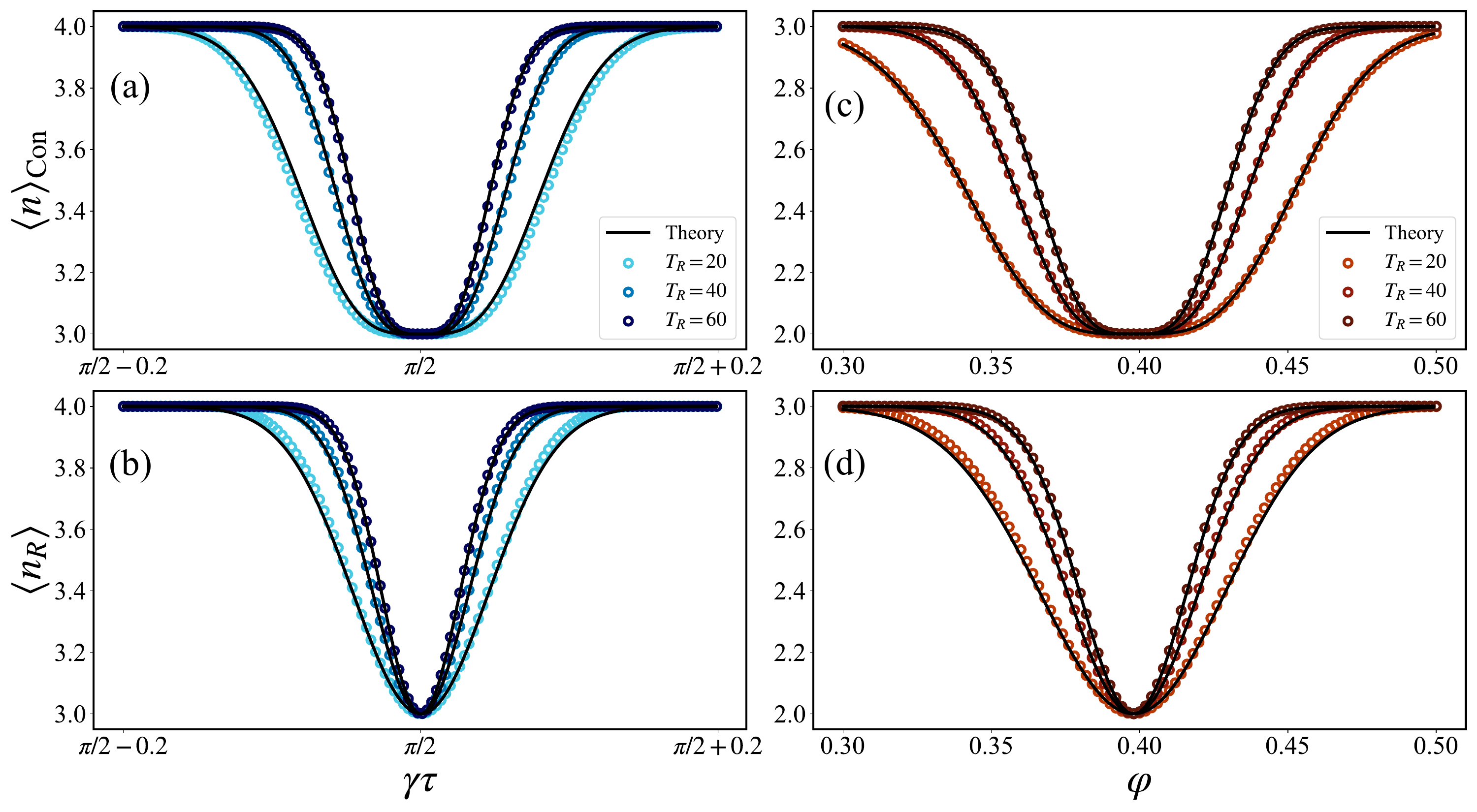}
    \end{center}
    \caption{(a) The conditional mean $\expval{n}_\text{Con}$ and (b) the restart mean $\expval{n_R}$
    as a function of $\gamma\tau$. 
    The model here is the benzene-type ring (Eq. (\ref{eq1}) with $L=6$), 
    and we work in the vicinity of its critical value $\gamma \tau = \pi/2$,
    with the transition $\langle n \rangle =4$ to $\langle n \rangle =3$.
    The black lines represent the theory (Eq. (\ref{mean6}) in (a) and Eq. (\ref{eq30}) in (b)).
    (c) The conditional mean $\expval{n}_\text{Con}$ and (d) the restart mean $\expval{n_R}$ 
    as a function of the magnetic flux $\varphi$, in the vicinity of a critical $\varphi$ around $0.4$.
    The model here is a three-site ring with a magnetic field Eq. (\ref{eq1m}), and $\gamma\tau$ is set as $3$. Here the transition is from $3$ to $2$.
    The theory (black lines) is plotted with (c) Eq. (\ref{eq33}) or (d) Eq. (\ref{eq34}).
    In all figures, from the bottom to the top line, 
    the restart time $T_R$ is $20$, $40$, and $60$, respectively.
    And the transition is narrowed when $T_R$ grows.
    The exact results represented by dots are obtained 
    with Eq. (\ref{renew}) and Eq. (\ref{cm}) or Eq. (\ref{resmean}).
    }
    \label{fig:comb}
    \end{figure*}

\section{Examples}\label{sec7}
{In this section,
to further demonstrate the universality of our theory, 
we consider a larger graph model, 
i.e. a ring of size $L=6$ 
as illustrated in Ref. \cite{yin2024restart}.
Together with the previously examined three-site ring model, 
this system exhibits resonance broadening when $\gamma\tau$ is tuned.
This naturally raises the question of whether our theory remains valid 
when other control parameters are varied. 
In light of this, 
we introduce an external magnetic field applied perpendicular to the graph plane. 
This modification gives rise to a chiral quantum walk
\cite{Zimboras2013,Matteo2023avs,Matteo2024enhanced}, 
where the energy spectrum depends additionally on the magnetic flux, 
thereby providing an alternative control parameter 
for tuning the resonance of energy phases.
As mentioned before, the existence of magnetic field breaks the time-reversal symmetry,
thus the {restart uncertainty relation} will be further validated in the aspect of universality.
}

\subsection{Benzene-type ring}
The first example is the benzene-type ring model governed by the Hamiltonian Eq. (\ref{eq1}) with $L=6$.
The energy levels are given by Eq. (\ref{eigenvalues1}),
namely, the distinct energies are 
$\{2\gamma,-2\gamma,\gamma,-\gamma\}$ with degeneracies $\{1,1,2,2\}$, respectively.
And the energy eigenstates are $\ket{E_k} = (1,e^{i\theta_k},e^{i2\theta_k},e^{i3\theta_k},e^{i4\theta_k},e^{i5\theta_k})^T/\sqrt{6}$ 
with $\theta_k = {\pi k / 3}$, and $T$ for matrix transpose. 
Hence the overlaps $p_k$ corresponding to distinct energies are 
$p_{\pm 2\gamma} = 1/6$, and $p_{\pm \gamma} = 1/3$, 
with the subscript standing for the energy values.
We therefore expect that except for a small subset of $\gamma\tau$'s, $\expval{n} = 4$. 
When $\gamma\tau={(2m+1)\pi/ 2}$ with $m=0,1,2,\dots$,
the mean $\expval{n}$ jumps from $4$ to $3$, 
since the energy phase factors $\{e^{i2\gamma\tau},e^{-i2\gamma\tau}\}$ are merging on the unit circle (one can readily check that the other two phase factors do not match). 
We will choose the resonance at $\gamma \tau = 3 \pi/2$ in the following.
Straightforward calculation using Eq. (\ref{condmean}) gives
%
%
%
\begin{equation}\label{mean6}
\begin{aligned}
    \langle n \rangle_{\text{Con}} 
    &= 4 - \left(1 +  T_R \xi^2 \right)
      \exp \left(- T_R \xi^2 \right) , \\
    &\text{with } \xi = \sqrt{3} \left( 4\gamma\tau\; \text{mod}\; 2\pi \right)/2 .
\end{aligned}
\end{equation}
%
Clearly this formula exhibits the transition from $4$ to $3$ and back as we vary $\gamma\tau$.
Using Eq.~(\ref{rmen}), the mean of non-constrained hitting time $n_R$ is
\begin{equation}\label{eq30}
    \langle n_R \rangle 
    = 4 - 
      \exp \left(- T_R \xi^2 \right).
\end{equation}
See Fig.~\ref{fig:comb}(a)(b) for numerics, 
where an excellent agreement between the theory and exact results is witnessed.
From lower to upper curves, the restart time $T_R$ is correspondingly $20$, $40$ and $60$,
with the full width at half maximum being smaller as $T_R$ grows.

{\bf Remark}: 
Exact results can be obtained as follows. 
One solves Eq.~(\ref{Fn}) 
to find the probability of the first detection $F_n$.
With definitions of the conditional mean Eq. (\ref{cm}), 
and restarted mean Eq. (\ref{resmean}), 
one may use a simple program like {\em Mathematica} to generate numerically exact results.

\subsection{Ring with magnetic flux}
In previous examples, 
we tuned $\gamma\tau$ to manipulate the matching of energy phase factors $e^{- iE_k\tau}$, 
and now we will change the energies themselves.
Here we choose to introduce an external magnetic field to the three-site ring model, 
the resulting magnetic flux $\varphi$.
Specifically, the tight-binding Hamiltonian we used before now reads 
\cite{Zimboras2013,Matteo2023avs,Matteo2024enhanced}
%
\begin{equation}
\label{eq1m}
H = -\gamma \sum_{x=0}^{2} \left(e^{i \varphi } \ket{x} \bra{x+1}+ e^{-i \varphi } \ket{x} \bra{x-1} \right),
\end{equation}
%
and here $L=3$ so we consider a three-site ring.
The time-reversal symmetry is now broken.
The eigenvalues of the Hamiltonian are
\begin{equation}
\label{eigenvalues2}
    E_k =   - 2 \gamma 
            \cos
            \left( 
            {2 \pi k \over 3} + \varphi 
            \right),  
            \ k=0,1,2 \ .
\end{equation}
For generic flux the degeneracy of the energy levels found for the case $\varphi=0$ is removed,
hence typically, i.e. for most $\gamma\tau$, and the initial/target state set as one node on the graph, 
we expect if $T_R\to \infty$ to find $\langle n \rangle =3$. 
Given $\varphi\in[0.3,0.5]$ and $\gamma\tau=3$, 
we encounter the case where the absolute value of a single eigenvalue of the survival operator ${\cal S}$ approaches 1 
when $\varphi$ is near 0.3979,
with the energy phase factors $\{ e^{- iE_0\tau},e^{- iE_2\tau} \}$ merging 
(the corresponding overlaps are $p_0=p_2=1/3$). 
With Eq. (\ref{condmean}), the conditional mean for $T_R$ measurement attempts is
\begin{equation}\label{eq33}
\begin{aligned}
    \expval{n}_\text{Con}   &= 3 - \left( 1+ T_R \zeta^2 \right) 
                            \exp \left( - T_R \zeta^2 \right), \\
    &\text{with } 
    \zeta = 
    \sqrt{6} 
    \left[ 2\sqrt{3} \gamma\tau \sin(2\pi/3 + \varphi) \; \text{mod}\; 2\pi \right] 
    \big/ 4.
\end{aligned}
\end{equation}
%
Similarly, with Eq.~(\ref{rmen}), we have for the same model,
\begin{equation}\label{eq34}
    \expval{n_R} = 3 - \exp \left( - T_R \zeta^2 \right),
\end{equation}
%
In Fig. \ref{fig:comb}(c)(d), we plot the theory in comparison with the exact results,
with $T_R$ chosen as $20, 40$ and $60$ from lower to upper curves.
We see that the width of the transition is becoming narrower as we increase $T_R$, 
and that the theory perfectly matches this $w=3\to w=2$ transition.

\begin{figure*}[tp]
\begin{center}
\includegraphics[width=0.90\linewidth]{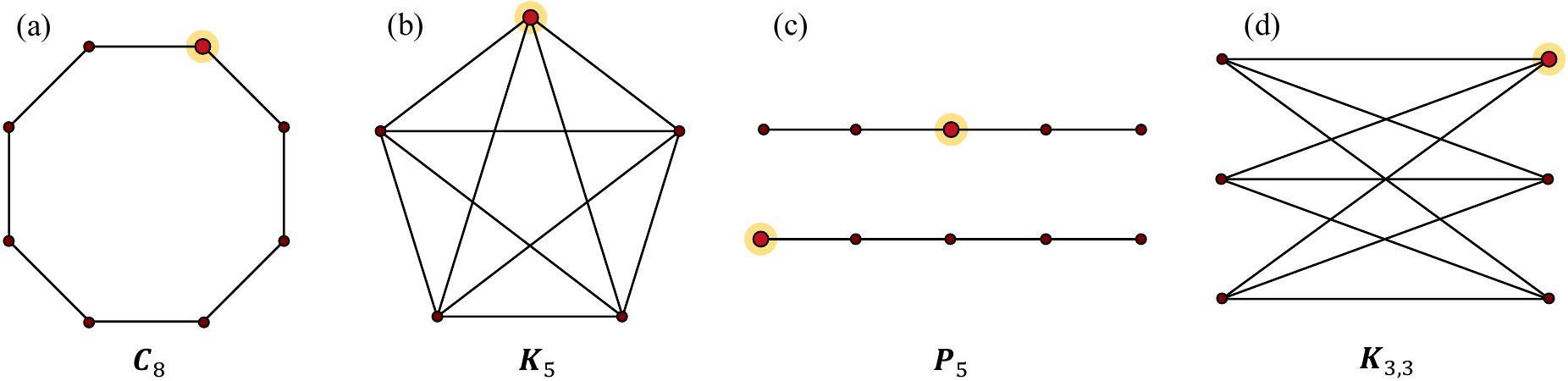}
\end{center}
\caption{
Schematics for the graphs under investigation.
From the left to the right are the examples:
(a) ring of size $L$, $C_L$, in the figure $L=8$,
(b) complete graph of size $L$, $K_L$, in the figure $L=5$,
(c) finite segment of size $L$, $P_L$, in the figure $L=5$,
(d) complete bipartite graph $K_{m,n}$, in the figure $m=n=3$, and the size is $L = m+n = 6$.
We will verify our theory using different sizes.
For graph (c), the target site will be chosen at the end or the mid
(marked with larger red vertices),
which leads to non-identical resonance widths. 
{Recall that the adjacency matrix of these graphs, 
defined the tight-binding $H$, from which the unitary is obtained.}
}
\label{fig:examgraphs}
\end{figure*}
\begin{table*}[t]
\centering
\caption{
The winding number $w$, 
the maximal difference between energies 
$\Delta E_{\rm m} = E_{\rm max} - E_{\rm min}$,
used in $\widetilde{\Delta E \tau} = (\tau \Delta E_{\rm m} \mod 2 \pi)$,
and the parameter $\lambda = p_+ p_- / (p_+ + p_-)^3$,
for different graphs with $L$ vertices, 
including even rings $C_L$, complete graphs $K_L$,
finite segments $P_L$, and complete bipartite graphs $K_{L/2,L/2}$.
Only for the segment, i.e. the $P_L$ graph, 
the location of the target, denoted $x_{\rm d}$, 
is important.
For other graphs, the translational invariance guarantees that the choice of $x_{\rm d}$
does not influence the results here.}
\label{uncTab}
\sisetup{detect-weight,mode=text}
\renewrobustcmd{\bfseries}{\fontseries{b}\selectfont}
\renewrobustcmd{\boldmath}{}
\newrobustcmd{\B}{\bfseries}
%
\begin{tabular}{ cccccccccccccccccccc } 
 \Xhline{2\arrayrulewidth}
 \multicolumn{4}{c}{ \B \,\,\,Graph\,\,\,} & 
 \multicolumn{4}{c}{\makecell{ ${\bm C_L}$, $L$ is even}} & 
 \multicolumn{4}{c}{\makecell{ ${\bm K_L}$}} & 
 \multicolumn{4}{c}{\makecell{ ${\bm P_L}$, $L$ is odd}} & 
 \multicolumn{4}{c}{\makecell{ ${\bm K_{{L\over2},{L\over2}}}$}}  \\~\\
 \multicolumn{4}{c}{${\bm w}$} & 
 \multicolumn{4}{c}{\,\,${L / 2 +1}$\,\,}  & 
 \multicolumn{4}{c}{\,\,$2$\,\,}  & 
 \multicolumn{4}{c}{\makecell{$L$, $x_{\rm d}=1$; \\
 ${(L+1) / 2}$, $x_{\rm d} = {(L+1) / 2}$}}  & 
 \multicolumn{4}{c}{\,\,$3$\,\,} \\~\\
 \multicolumn{4}{c}{${\bm \Delta E_{\rm m}}$} & 
 \multicolumn{4}{c}{$4\gamma$} & 
 \multicolumn{4}{c}{$\gamma L$} & 
 \multicolumn{4}{c}{$4\gamma \cos\left[ {\pi / (L+1)} \right]$} & 
 \multicolumn{4}{c}{$\gamma L$} \\~\\
 %
 \multicolumn{4}{c}{${\bm \lambda}$} & 
 \multicolumn{4}{c}{${L / 8}$} & 
 \multicolumn{4}{c}{${(L-1) / L^2}$} & 
 \multicolumn{4}{c}{\makecell{${(L+1)^3 / 16\pi^2}$, $x_{\rm d}=1$; \\
 ${(L+1) / 16}$, $x_{\rm d}={(L+1) / 2}$}} & 
 \multicolumn{4}{c}{${L / 8}$} \\
 %
 \Xhline{2\arrayrulewidth}
\end{tabular}
\end{table*}
%
%

\section{System size effects}
\label{SizeEffects}

An issue that naturally arises is the effect of system size on our main results. 
The energy spacing and eigenstate property, which may relate to system size, 
are captured by the factors $\widetilde{\Delta E \tau}$ and $\lambda$
in Eqs. (\ref{condmean}) and (\ref{rmen}).
For the completeness of interest, 
we now discuss the relation between the restart uncertainty relation 
and the size of the system. 

We will only focus on the restarted mean $\expval{n_R}$ below,
since similar behaviors are found for the conditional mean $\expval{n}_\text{Con}$.
And for the models, we will investigate the following graphs:
(a) ring of size $L$, $C_L$, 
(b) finite segment of size $L$, $P_L$,
(c) complete graph of size $L$, $K_L$, 
(d) complete bipartite graph $K_{m,n}$ with $L=m+n$.
%
%
%
Without delving into details, we have summarized in Table \ref{uncTab}, 
the values of parameters in Eqs. (\ref{condmean}) and (\ref{rmen}),
for different graphs, 
with the resonance chosen at 
$\exp (- i E_{\rm \max} \tau) \sim\exp ( - i E_{\rm min} \tau)$, 
{where $E_{\rm \max}$ and $E_{\rm \min}$ are the maximum and minimum of energies 
of the system respectively}.
It is clearly shown that different graph structures lead to various relations
between the width of transitions and the system size $L$.

%
\begin{figure*}[t]
\begin{center}
\includegraphics[width=0.85\linewidth]{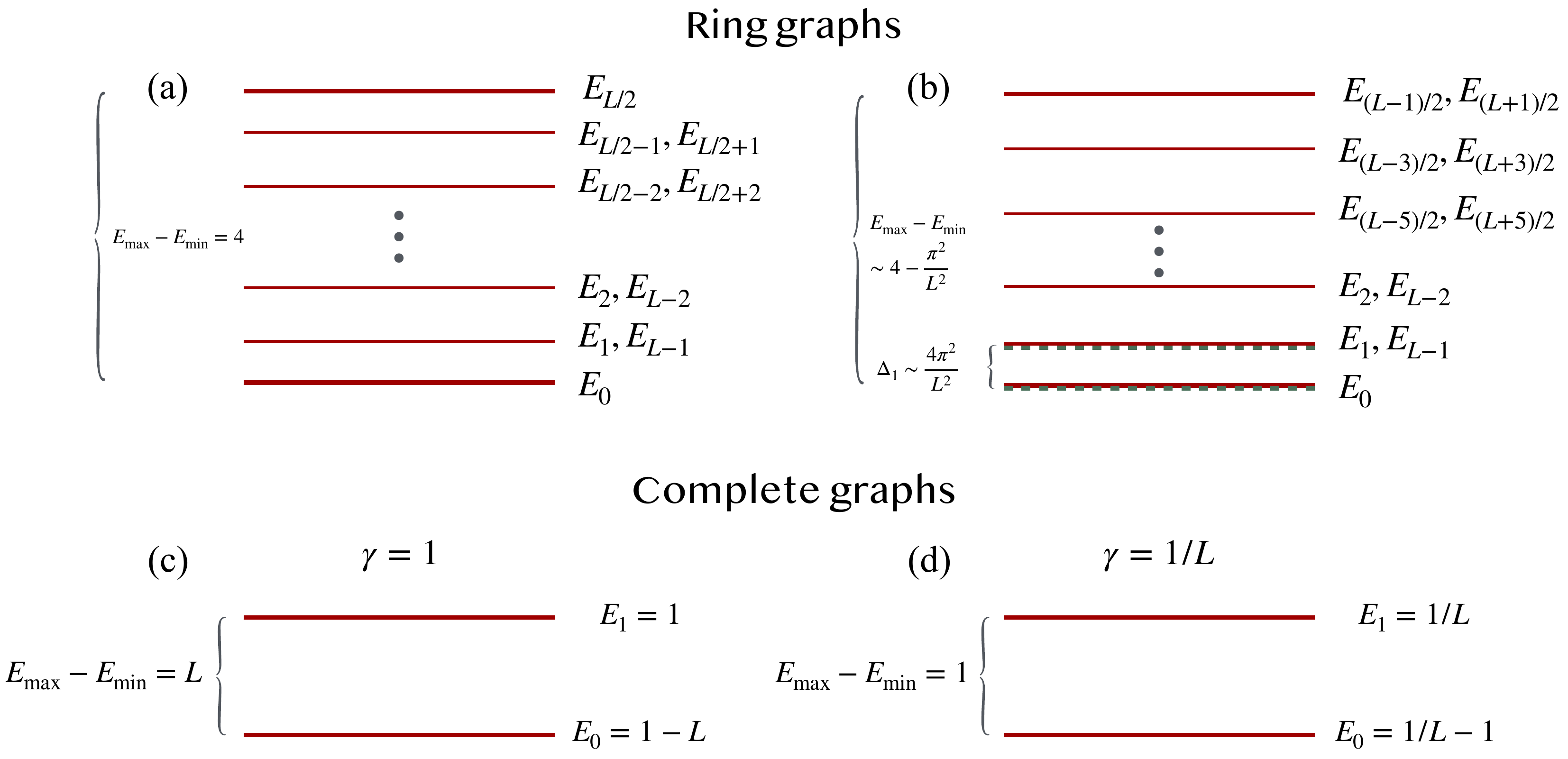}
\end{center}
\caption{
The energy levels of ring graphs and complete graphs. 
In (a) we present the case of even $L$, while in (b) $L$ is odd.
We consider the resonance related to the largest energy
and the lowest energy (ground state energy),
which we called the min-max condition. 
{The energy gap is then denoted $\Delta E_\text{m}$.}
As a second option
we choose the ground state energy and the first excited state energy.
The dispersion relation for rings is $E_k = -2 \gamma \cos(2\pi k /L)$ 
with $k=0,1,2,\dots,L-1$ and $\gamma=1$.
Here $\gamma$ is the hopping amplitude between nodes, 
namely $H$ is the adjacency matrix of the graph multiplied by $\gamma$.
For complete graphs (subplot (c)), the energies are $1$ and $1-L$.
As typically used in literature, 
the hopping rate $\gamma$ is set as inversely proportional 
to the number of edges of each vertex, 
see subplot (d) where the energy difference is $E_{\rm max}- E_{\rm min} = 1$.
}
\label{fig:enlring}
\end{figure*}
%
\subsection{Ring models} 
We start with the ring model (Fig. \ref{fig:examgraphs}(a)),
which is used for demonstration purposes in the manuscript. 
{Here due to symmetry, all localised initial conditions are identical, 
for the recurrence time problem under study.}
As mentioned, energies of the ring model of size $L$ are 
$E_k = - 2 \gamma \cos \theta_k$ 
with $\theta_k = {2 \pi k / L}$ and $k=0,1,2,\dots, L-1$, 
and overlaps are $\left| \braket{x}{E_k}\right|^2 = 1/L$ for any node $x$ , 
the broadening can be easily associated with the system size $L$ (assuming even $L$).
See Figs. \ref{fig:enlring}(a) and (b) for schematics of its energy structures,
where the parity of $L$ plays a role.
We start the discussion where the pair of energies is 
$E_{\rm max}$ and $E_{\rm min}$, whose phase factors match,
and then consider the case when we chose the energy difference 
between the ground state and the first excited state 
(this is based on odd rings, otherwise the transition will be $w \to w - 2$ 
which is left for future study). 

\begin{figure}[ht]
\begin{center}
\includegraphics[width=0.95\linewidth]{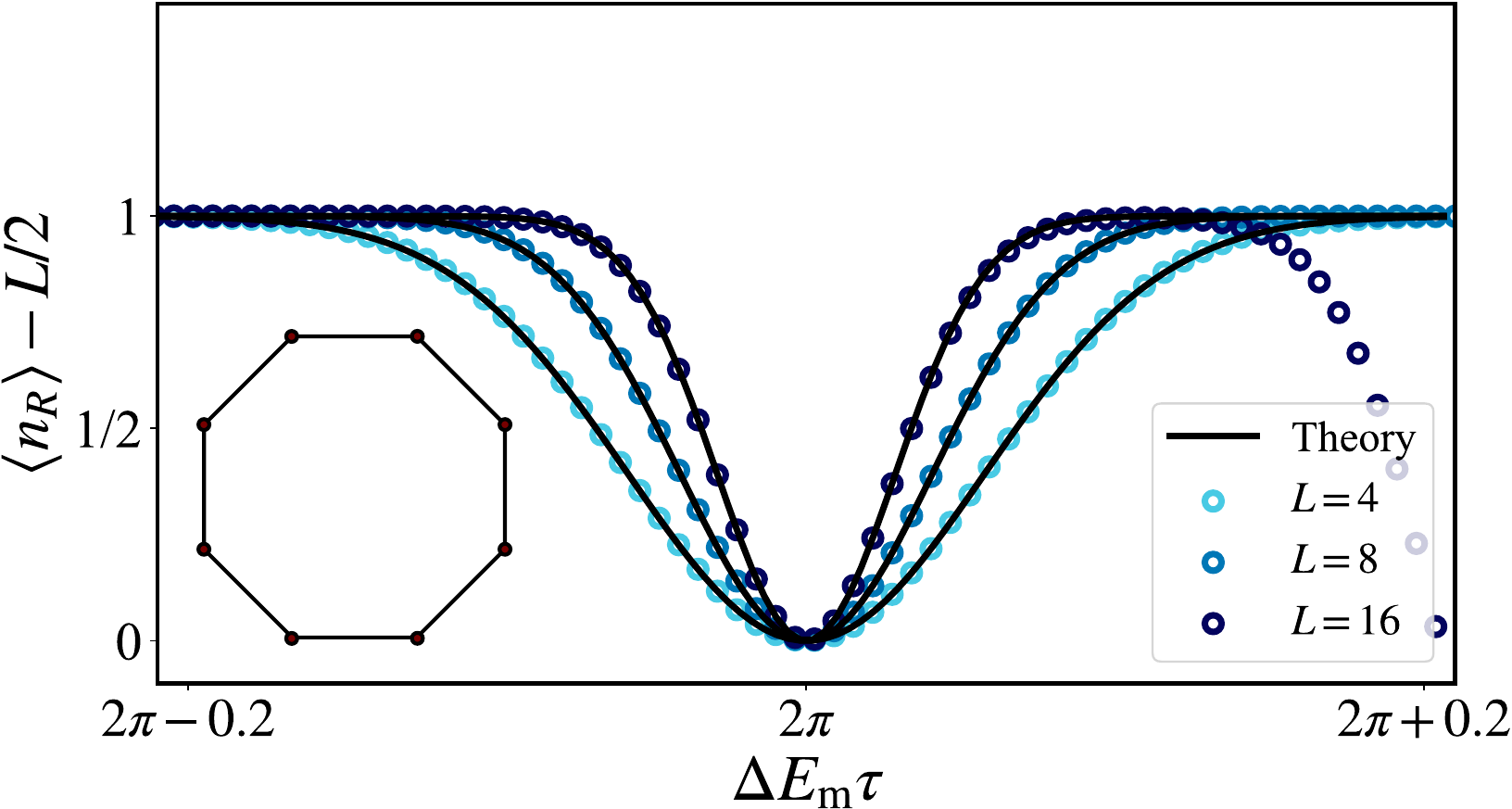}
\end{center}
\caption{Restarted mean hitting times 
versus $(E_{\rm max} - E_{\rm min}) \tau$, 
for ring graphs of various sizes $L$ (see inset for an example).
The resonances become narrower as we increase the size of the system. 
Recall, that difference between the largest and ground-state energies, 
$E_{\rm max} - E_{\rm min} = 4\gamma$, is size-independent, and we choose $\gamma=1$. 
We shift the mean by $L/2$, to focus on the width of the transition. 
The numerical results are obtained with Eqs. (\ref{Fn}, \ref{cm}, \ref{resmean}), 
and this perfectly matches our theory, 
see Eq. (\ref{eq4}).
The deviation on the right for $L=16$ is caused by the proximity of another resonance.
$T_R=300$ is used here. 
Similar results for $\expval{n}_\text{Con}$ were also tested, 
and not presented hereinafter.
}
\label{fig:ring0}
\end{figure}

For the resonance between the ground state and the highest energy state (denoted min-max choice), 
when $L$ is even,
the phase factors $\{e^{-i2\gamma\tau}, e^{i2\gamma\tau} \}$ merge,
we have $\widetilde{\Delta E \tau } = \tau \Delta E_{\rm m} \mod 2\pi 
= 4\gamma\tau \mod 2\pi$
(and now we set $\gamma$ as $1$),
and $\Delta E_{\rm m} = E_{\rm max} - E_{\rm min} = 4\gamma$.
Thus, for even $L$,

%
\begin{equation}
    \begin{split}
        \langle n_R\rangle 
        &= 
        w
        - 
        \exp[- L(\widetilde{\Delta E \tau })^2  T_R /8 ], 
    \end{split}
    \label{eq4}
\end{equation}
where $w=(2 + L)/2$.
We note that for odd ring, $w=(L+1)/2$, 
and $\Delta E_{\rm m}$ is $4-\pi^2/L^2$.
See Fig. \ref{fig:enlring}(b).
Thus, with $L$ increasing, the broadening of the transition will be narrower, 
for all the rings with odd or even number of nodes.
See Fig. \ref{fig:ring0}, where we present numerical confirmation for even rings.
%

%
%

However, if we consider the resonance related to
the ground state and the first excited state, for the odd rings, 
which leads to the transition $w\to w-1$,
the $L$ dependence of the energy difference will be distinct.
In this case it follows that
$\widetilde{\Delta E \tau } = \tau \Delta_1 \mod 2\pi 
= (E_{\rm 1st} - E_{\rm g})\tau \mod 2\pi$
with $E_{\rm 1st} - E_{\rm g} \sim 1/L^2$,
i.e. the energy difference shrinks when the system size $L$ grows 
(see Fig. \ref{fig:enlring}(b)).
The parameter $\lambda$ is still proportional to $L$,
and then the term $\lambda (\widetilde{\Delta E \tau })^2$ 
is proportional to $1/L^3$, when $\tau$ is tuned close to the resonance.
Hence this will result in an increasing width of the resonance
as we increase the size $L$.

{Another important quantity affected by system size 
is the detection probability within $T_R$ measurements 
(Eq. (\ref{pdet15})), 
which for the ring model, under the min-max choice, reads:}
{
\begin{equation}\label{pdetring}
    P_\text{det} (T_R) 
    \sim 
    1- {L (\widetilde{\Delta E \tau})^2 \over 8}
    \exp \left[- {L (\widetilde{\Delta E \tau})^2 \over 8} T_R \right],
\end{equation}
%
where we substitute 
\begin{equation}\label{eqpdet1}
    \epsilon^2 = {L (\widetilde{\Delta E \tau})^2\over 8}
\end{equation}
into Eq. (\ref{pdet15}).
From Eq. (\ref{pdetring}) we can see that, 
$P_\text{det}(T_R)\to 1$ 
when $\widetilde{\Delta E \tau}$ goes to $0$ 
or $\widetilde{\Delta E \tau} \gg 1$.
Hence $P_\text{det}(T_R)$ versus $\tau$ will exhibit a $W$-like structure, similar to the behavior shown in Fig.~\ref{fig:Tridip}(d),
with its scaling with $L$ determined by $\lambda (\widetilde{\Delta E \tau })^2$ as a function of $L$. 
In Fig. \ref{pdetTR}, 
we provide validation of our theory Eq. (\ref{pdet15})
for various graph models (see below),
by presenting the rescaled $P_\text{det} (T_R)$ that is linear with $T_R$,
namely $-\ln\left[{1 - P_\text{det}(T_R)}\right] \sim T_R \epsilon^2 -2\ln \epsilon$.}


Now it is readily realized that
the system size $L$ has various ways of entering the expressions 
for energy levels and eigenstates. 
In the context of quantum walks on graphs, 
this means that the graphs, on which we dispatch quantum walkers, matter.
Different graph structures lead to different dispersion relations $E_k$, 
as well as the corresponding eigenvectors $\ket{E_k}$.
To explore how $L$ determines $\lambda$ and $\widetilde{\Delta E \tau}$, 
we checked other graphs.

\begin{figure}[tp]
    \begin{center}
    \includegraphics[width=\linewidth]{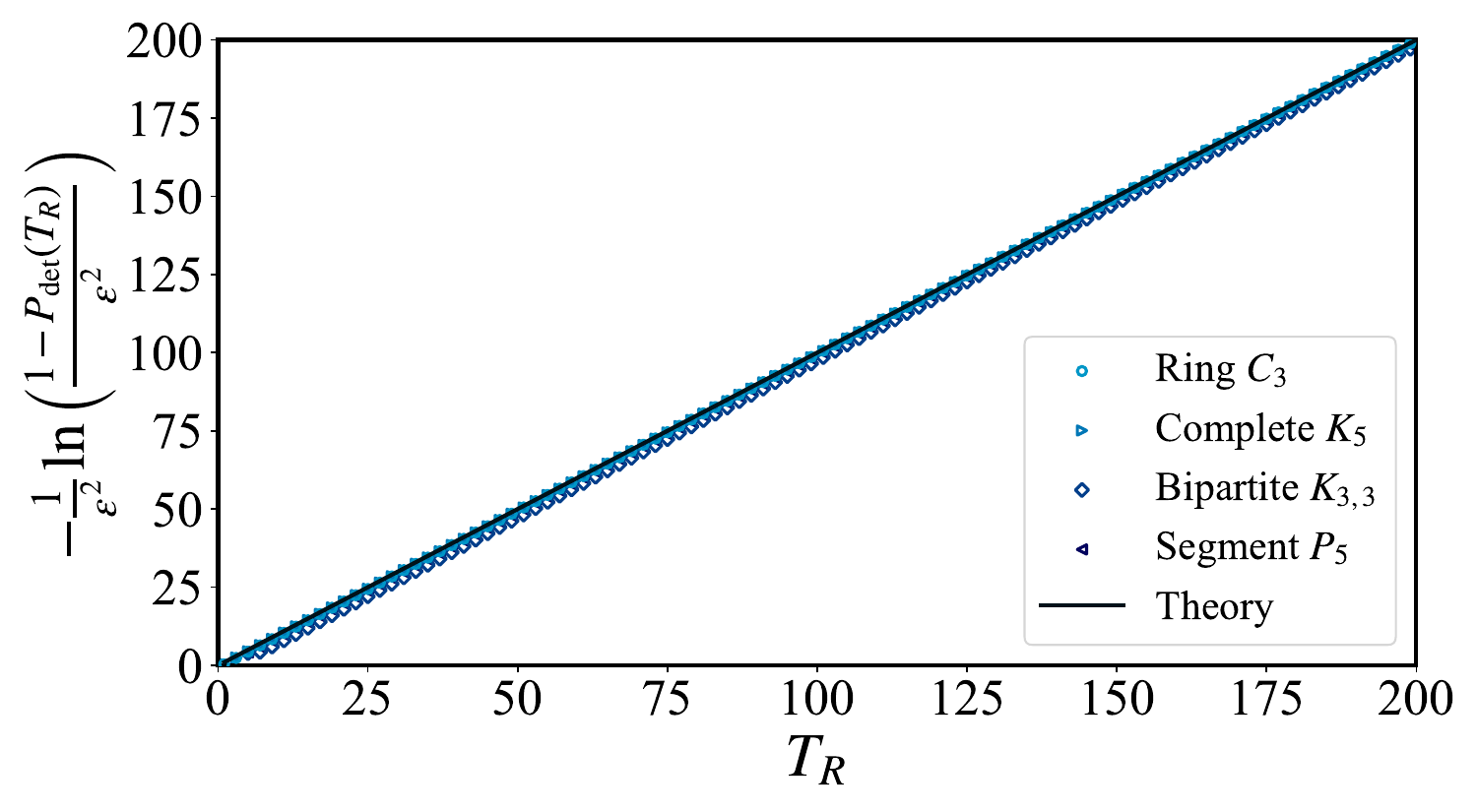}
    \end{center}
    \caption{
        From the theoretical expression for the detection probability $P_\text{det}(T_R)$, given by Eq. (\ref{pdet15}), we derive the general relationship:
        $-\frac{1}{\epsilon^2}\ln\left[\frac{1 - P_\text{det}(T_R)}{\epsilon^2}\right] \sim T_R$, 
        which is graph-independent and represented by the solid black line in the figure. 
        We use Eqs.~(\ref{eqpdet1},\ref{eqpdet2},\ref{eqpdet3},\ref{eqpdet4},\ref{eqpdet5}) to obtain $\epsilon^2$, 
        for specific graphs depicted in Fig.~\ref{fig:examgraphs}. 
        Exact results computed with Eq.~(\ref{Fn}) 
        are shown as colored symbols (circles, triangles, etc.). 
        The close agreement between theory and exact numerical results confirms the convergence of $P_\text{det}(T_R)$ and validates our theoretical approach.
    }
    \label{pdetTR}
\end{figure}
\begin{figure}[ht]
\begin{center}
\includegraphics[width=0.95\linewidth]{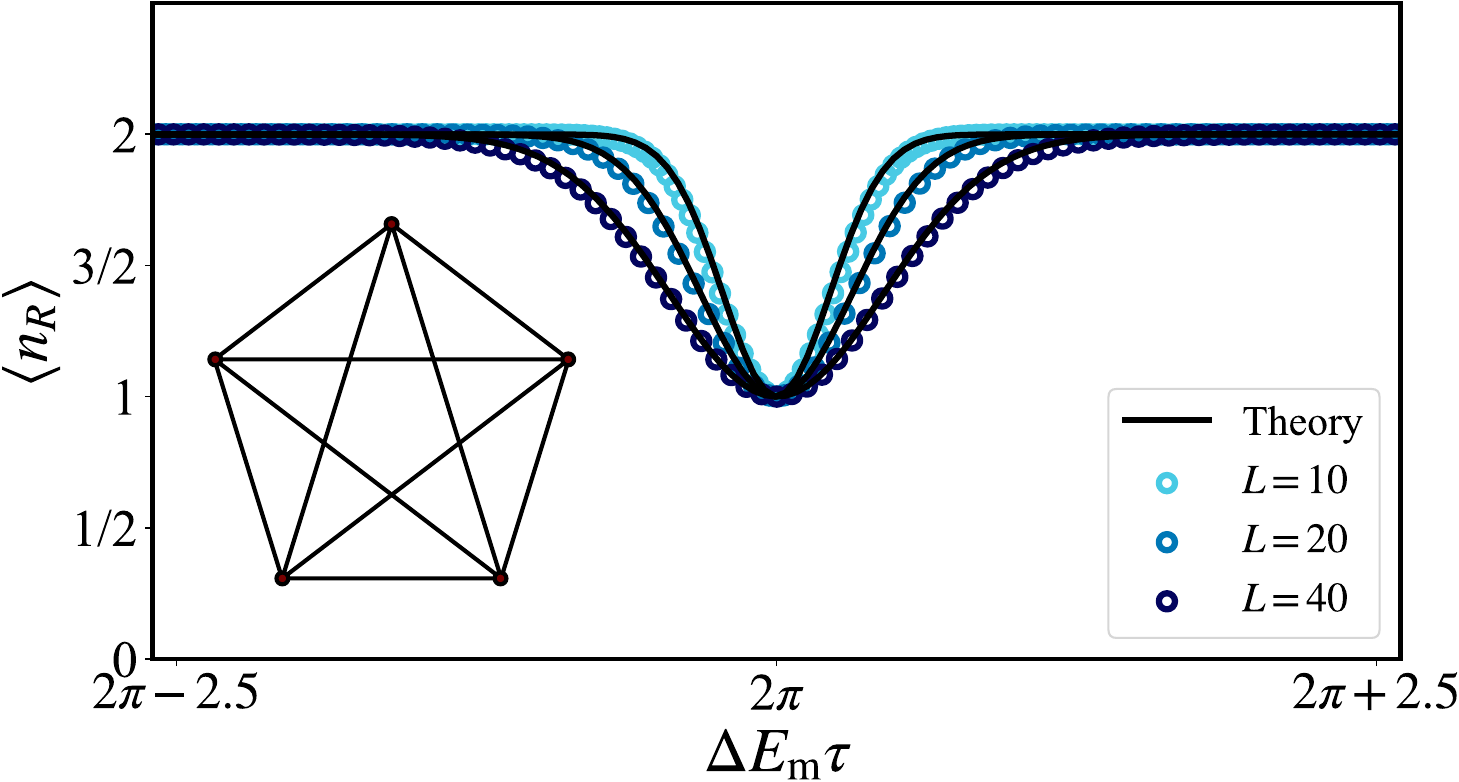}
\end{center}
\caption{
Restarted mean hitting time 
for complete graphs of different sizes 
(see inset for an example). 
The numerical results are obtained with Eqs. (\ref{Fn}, \ref{cm}, \ref{resmean}) in the main text, 
and the theory is computed with Eq. (\ref{eq5}).
Here $\Delta E_{\rm m} = 1$ and we used $\gamma=1/L$ for a fair comparison. 
Unlike Fig. \ref{fig:ring0}, the broadening becomes wider as the system size $L$ increases.
Here we used $T_R=100$.
}
\label{fig:cg1}
\end{figure}
%

\subsection{Complete graph models} 
One example is the complete graph, in which each vertex is connected to every other vertex. 
See Fig. \ref{fig:examgraphs}(b). 
Specifically, the governing Hamiltonian in matrix form, 
has all elements equal $-\gamma$ except for the diagonal. 
To achieve a fair comparison, the hopping rate is usually chosen as $\gamma=\gamma_0/L$, 
and we set $\gamma_0=1$ here.
There are merely two energy levels, $E_0=\gamma(1-L)$ and $E_1=\gamma$, 
and the eigenstate corresponding to $\gamma(1-L)$, 
or the ground state is $(1,1,1,\dots,1)/\sqrt{L}$, 
hence the overlaps, for any initial/target state, are $p_+ = 1/L$ and $p_- = (L-1)/L$. 
This further leads to $\lambda = p_+ p_- /(p_+ + p_-)^3 =(L-1)/L^2 \sim 1/L$ 
as $L$ is large.
Hence with Eq. (\ref{pdet15})
we have for large $L$,
%
\begin{equation}\label{pdetcg}
    1-P_\text{det}(T_R) \sim 
    [(\widetilde{\Delta E \tau })^2 / L] 
    \exp \left[ -(\widetilde{\Delta E \tau })^2 T_R / L \right],
\end{equation}
where $\widetilde{\Delta E \tau} 
= \tau \Delta E_{\rm m} \mod  2 \pi = L\gamma \tau \mod  2 \pi$,
and 
\begin{equation}\label{eqpdet2}
    \epsilon^2 = {(\widetilde{\Delta E \tau })^2 \over L}
\end{equation}
is substituted into Eq. (\ref{pdet15}).
Since when $\widetilde{\Delta E \tau}$ goes to or deviates away from $0$,
$1-P_\text{det}(T_R)$ approaches $0$,
$P_\text{det}(T_R)$ is a $W$-shape function
as $\gamma\tau$ is varied around the resonance at $\widetilde{\Delta E \tau}=0$.
We numerically confirm Eq. (\ref{pdetcg}) in Fig. \ref{pdetTR}.
%
%
Using Eq. (\ref{rmen}), we obtain
\begin{equation}
    \begin{split}
        \langle n_R \rangle &\sim 
        w
        - 
        \exp[- (\widetilde{\Delta E \tau })^2 T_R/L ], 
    \end{split}
    \label{eq5}
\end{equation}
where $w=2$. 
From here we see that if choosing $\gamma$ independent on $L$, say $\gamma=1$,
we will have $\widetilde{\Delta E \tau} = L \tau \mod  2 \pi$ 
since the energy difference becomes $L$,
as shown in Fig. \ref{fig:enlring}(c).
Then Eq. (\ref{eq5}) becomes 
$\langle n_R \rangle \sim w - \exp[- L(\tau - 2\pi/L)^2 T_R ]$ when $\tau\simeq 2\pi/L$,
indicating again a decreasing width of resonance as $L$ grows.

As mentioned above, 
we could also choose $\gamma=1/L$ as done in the literature of quantum walks, 
which leads the energy difference to $\Delta E_{\rm m} =1$, 
as shown in Fig. \ref{fig:enlring}(d).
Then we have $\langle n_R \rangle \sim w - \exp[- (\tau \mod 2\pi)^2 T_R/L]$,
suggesting an increasing width of resonance with the system size $L$ increasing.
%
%
See Fig. \ref{fig:cg1} for the graphic demonstration. 

%
\begin{figure}[ht]
\begin{center}
\includegraphics[width=0.95\linewidth]{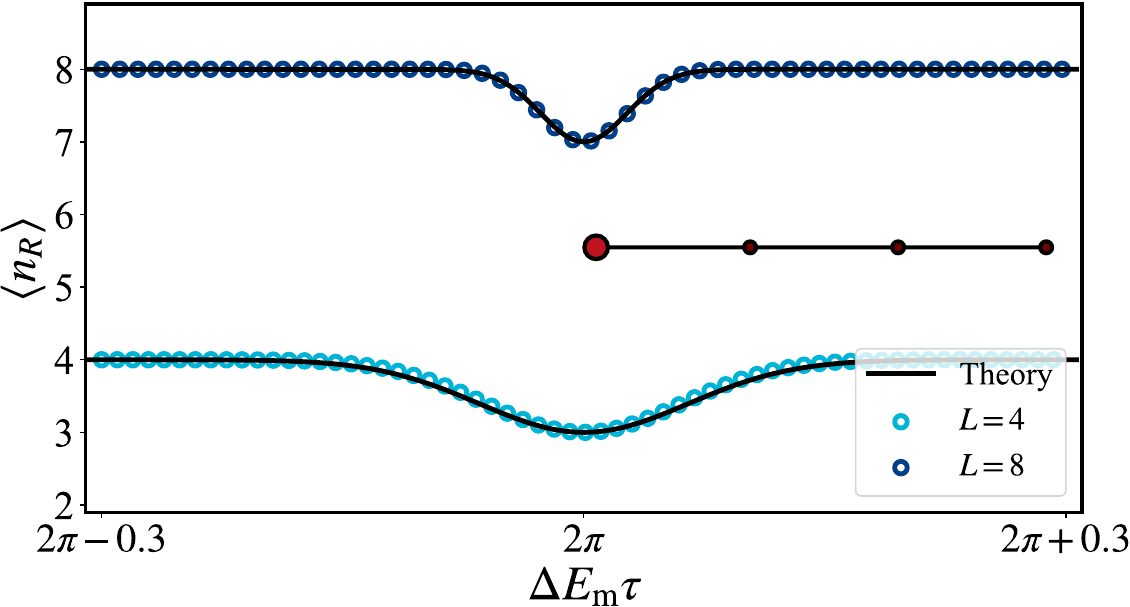}
\end{center}
\caption{
Restarted mean hitting time 
for finite segments of different sizes.
Repeated measurements are made on the leftmost node 
(see schematics in the inset, where the larger circle points to the measured node).
The numerical results are obtained with Eqs. (\ref{Fn}, \ref{cm}, \ref{resmean}) in the main text, 
and the theory is computed with Eq. (\ref{line1}).
We see that the broadening becomes narrower as the system size $L$ increases.
Here $T_R=150$. 
}
\label{fig:lineend}
\end{figure}
\begin{figure}[ht]
\begin{center}
\includegraphics[width=0.95\linewidth]{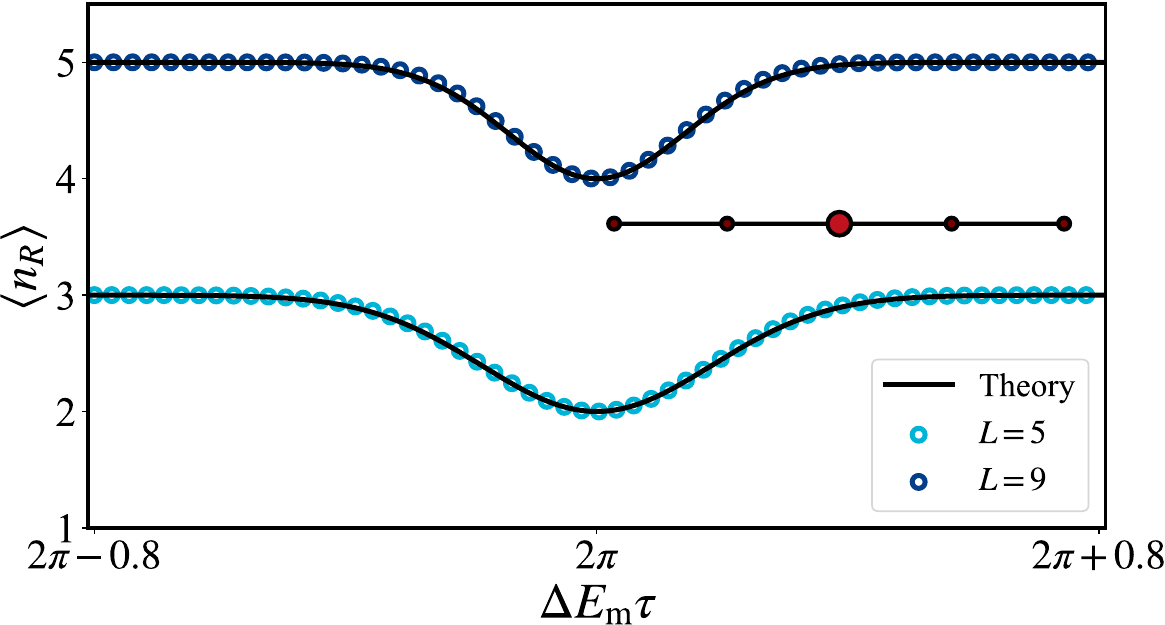}
\end{center}
\caption{
Restarted mean hitting time 
for finite segments of different sizes, 
here the target is set at $x_\text{d}=(L+1)/2$.
{This state is marked in red in the inset.}
The numerical results are obtained with Eqs. (\ref{Fn}, \ref{cm}, \ref{resmean}) in the main text, 
and the theory is computed with Eq. (\ref{line2}).
We see that the broadening becomes narrower as the system size $L$ increases.
$T_R=40$ is used.
Here measurement is performed on the middle node, see inset.
}
\label{fig:linemid}
\end{figure}
%
\subsection{Linear segments}
We also checked linear segments
(see Fig. \ref{fig:examgraphs}(c)). 
Around the resonance where 
$\{e^{-i\tau E_\text{max}}, e^{-i\tau E_\text{min}} \}$ merge, 
the prefactor of $(\tau-\tau_c)^2$ (with $\tau_c$ the resonance $\tau$) 
is proportional to $L^3T_R$, suggesting again narrower broadening as $L$ grows.
More concretely, for a line of size $L$, the Hamiltonian is 
$H = -\gamma \sum_{x=1}^{L-1} (\ket{x} \bra{x+1} + \ket{x+1} \bra{x})$,
whose energy levels are 
\begin{equation}
    E_k = -2\gamma \cos[k\pi/(L+1)], \,\, k=1,2,\dots, L,
\end{equation}
and the corresponding eigenvectors are
%
\begin{equation}
    \ket{E_k} = \sqrt{2/(L+1)} \sum_{x=1}^L \sin[k\pi x/(L+1)] \ket{x}.
\end{equation}
Hence there are $L$ distinct energies (no degeneracy), 
with the largest (lowest) energy $E_L=2\gamma \cos[\pi/(L+1)]$ ($E_1=-E_L$),
and the overlaps, for certain target site $x_\text{d}$, are
$p_k = |\bra{x_\text{d}}\ket{E_k}|^2 = [2/(L+1)] \sin^2[k\pi x_\text{d}/(L+1)]$.
Assuming odd $L$,
we consider $x_\text{d}$ at either end of the line, or the middle of the line,
namely, $x_\text{d}=1$ or $x_\text{d}=(L+1)/2$, leading to 
$p_k = |\bra{1}\ket{E_k}|^2 = [2/(L+1)] \sin^2[k\pi /(L+1)]$,
or
$p_k = |\bra{{L+1 \over 2}}\ket{E_k}|^2 = [2/(L+1)] \sin^2(k\pi /2) = [1-(-1)^k] / (L+1)$,
respectively.
We find that the $p_k$'s are non-zero for the former case, 
while for the latter, $x_\text{d}=(L+1)/2$, there appear $p_l=0$ when $l$ is even.
This leads to different winding numbers for the two choices of $x_\text{d}$, 
since $w$ is equal to the number of distinct phases $e^{-iE_k\tau}$ 
associated with non-zero $p_k$.
Hence $w=L$ for the case $x_\text{d}=1$,
and $w=(L+1)/2$ for the case $x_\text{d}=(L+1)/2$.
We now focus on the resonance 
where phases $\{ e^{-i\tau E_1}, e^{-i\tau E_L} \}$ merge at
$\tau_c=2\pi/|E_1-E_L| = \pi/2\gamma \cos[\pi/(L+1)]$,
which is the smallest resonance $\tau$ except for $\tau=0$.
For the target at one end of the line, $x_\text{d}=1$,
the corresponding overlaps to $E_1$ and $E_L$ are 
$p_1 = p_L = [2/(L+1)] \sin^2[\pi /(L+1)] \approx 2\pi^2/(L+1)^3$, 
with the approximation valid when $L$ is large.
For the case $x_\text{d}=(L+1)/2$, we have 
$p_1 = p_L = 2/(L+1)$.
Therefore, for $x_\text{d} =1$, namely the end node on the segment,
Eq. (\ref{pdet15}) for large $L$ becomes
\begin{equation}
\begin{aligned}
    P_\text{det} (T_R) \sim 1 - \epsilon^2_{\rm end} \exp(-T_R \epsilon^2_{\rm end}),
\end{aligned}
\end{equation}
where
\begin{equation}\label{eqpdet3}
\epsilon^2_{\rm end} = {(L+1)^3 \over 16\pi^2}
\left[ 4 \tau \cos({\pi / (L+1)})  \mod 2 \pi \right]^2.
\end{equation}
Then Eq. (\ref{rmen}) for large $L$ becomes
\begin{equation}\label{line1}
    \begin{split}
        \langle n_R \rangle &= 
        w - 
        \exp (-T_R \epsilon^2_{\rm end})
        , 
    \end{split}
\end{equation}
where $w = L$. 
And for the target site at the middle of the line, $x_\text{d} = (L+1)/2$, we have
%
\begin{equation}
    P_\text{det} (T_R) \sim 1 - \epsilon^2_{\rm mid} \exp(-T_R \epsilon^2_{\rm mid}), 
\end{equation}
where 
\begin{equation}\label{eqpdet4}
\epsilon^2_{\rm mid} = {L+1 \over 16} 
\left[ 4 \tau \cos({\pi / (L+1)})  \mod 2 \pi \right]^2.
\end{equation}
%
The restarted mean hitting time becomes 
\begin{equation}\label{line2}
    \begin{split}
        \langle n_R \rangle &= 
        w- 
        \exp (-T_R \epsilon^2_{\rm mid})
        , 
    \end{split}
\end{equation}
where $w=(L+1)/2$.
Clearly, 
these expressions exhibit a different dependence on system size.
See Figs. \ref{pdetTR}, \ref{fig:lineend} and \ref{fig:linemid} for numerical confirmation,
where the theory works well
and predicts the narrowing of broadening of resonances as the system becomes larger. 
%

%
\begin{figure}[t]
\begin{center}
\includegraphics[width=0.95\linewidth]{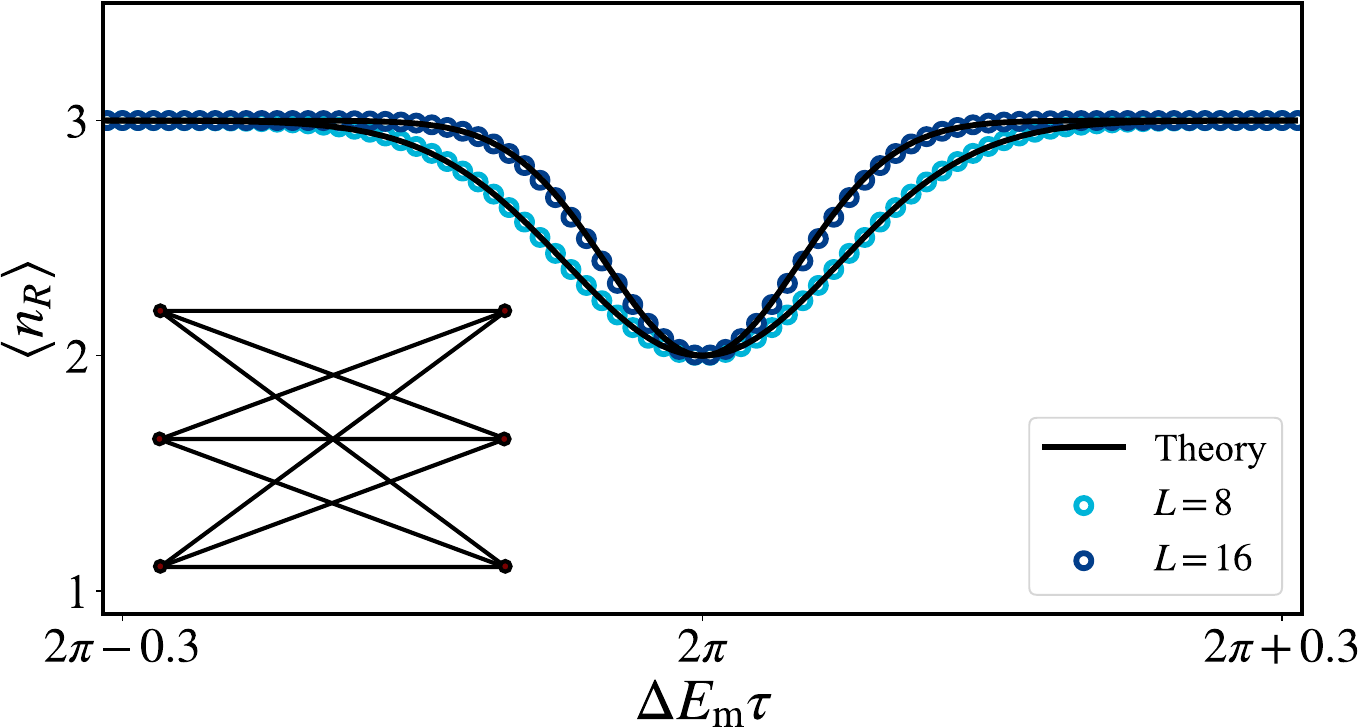}
\end{center}
\caption{
Restarted mean hitting time for complete bipartite graphs $K_{L/2,L/2}$ 
with different $L$ (see an example in Fig. \ref{fig:examgraphs}(d)).
The numerical results are obtained with Eqs. (\ref{Fn}, \ref{cm}, \ref{resmean}) in the main text, 
and the theory is computed with Eq. (\ref{cb}).
We see that the broadening becomes narrower as the system size $L$ increases.
Here $T_R=100$.
}
\label{fig:combip}
\end{figure}
%
\subsection{Bipartite graphs}
Another example is a complete bipartite graph, also called a complete bi-colored graph, 
usually denoted by $K_{l,m}$, see Fig. \ref{fig:examgraphs}(d). 
The vertices of the graph can be decomposed into two disjoint sets, 
containing $l$ and $m$ elements, respectively,
such that {\em no} two vertices within the {\em same} set are connected by an edge,
and every pair of vertices from {\em different} sets are connected.
See Fig. \ref{fig:examgraphs}(d) for schematics of $K_{3,3}$.
Here we use $K_{L/2,L/2}$ to demonstrate the influence of size $L$ 
on the restart uncertainty relation.
The Hamiltonian governing a quantum walk on such a graph is
$H = -\gamma 
\begin{bmatrix}
	O &C \\
	C &O
\end{bmatrix}
$
with $C$ a ${L\over 2} \times {L\over 2}$ matrix with all elements as $1$.
The energy levels are 
$\gamma \{ -L/2, 0, L/2 \}$.
The eigenvectors corresponding to the lowest and largest energies are
$\ket{E_0} = \left(-1, -1, -1, \cdots, -1,-1,-1, \cdots \right)^T/\sqrt{L}$,
and $\ket{E_2} = \left(-1, -1, -1, \cdots, 1,1,1, \cdots \right)^T/\sqrt{L}$.
This gives, around the resonance where $\{ e^{-i\tau E_0}, e^{-i\tau E_2} \}$ merge,
the overlaps $p_0 = p_2 = 1/L$, for any node as the target site.
Hence the parameters are straightforwardly calculated, namely
$\lambda = L/8$, and $\widetilde{\Delta E \tau} = \tau L \mod 2\pi$ 
($\gamma$ is set as $1$).
In the vicinity of the resonance $\widetilde{\Delta E \tau}\simeq 0$, 
$(\tau L \mod 2\pi)^2$ becomes $(\tau L - 2\pi)^2$.
Thus, the detection probability within $T_R$ measurement attempts is
%
\begin{equation}
    1 - P_\text{det}(T_R) \sim 
        \epsilon^2
        \exp
        \left(
        - \epsilon^2 T_R 
        \right),
\end{equation}
where 
\begin{equation}\label{eqpdet5}
    \epsilon^2 = {L^3 \over 8} \left(\tau - {2\pi\over L} \right)^2.
\end{equation} 
See Fig. \ref{pdetTR} for numerical confirmation.
The statistical measure of mean hitting time then becomes
%
%
%
%
\begin{equation}\label{cb}
    \begin{split}
        \langle n_R \rangle &= 
        w - 
        \exp
        \left(
        - \epsilon^2 T_R 
        \right), 
    \end{split}
\end{equation}
where $w = 3$.
In Fig. \ref{fig:combip} we present the numerical results 
calculated with Eqs. (\ref{Fn}, \ref{cm}, \ref{resmean}), 
and our theory agrees excellently with the numerics.
Therefore, as theoretically predicted and numerically seen,
the increasing system size leads to more abrupt transitions of the mean hitting times,
namely the resonance is narrowed as we increase $L$.

To summarize, the investigation of system size effects 
leads to the discovery of rich physics. 
The $L$ dependence of resonance broadening is not obeying a generic law,
but relies on the choice of resonance and the connectivity of models. 
Our focus on different energy levels and systems reveals varying behaviors,
such as broadening with increasing or decreasing system size.

\section{Discussion}
\label{sec9}
%
%
In this study following Ref. \cite{yin2024restart}, 
we investigate an {uncertainty-like relation} in quantum hitting times,
which is defined by repeated measurements, 
and currently measurable using quantum computers 
\cite{Wang2023,yin2024restart} with the approach of restart.
Through the exploration of the impact of system energies and the measurement period (sampling time) $\tau$ on this problem, 
we unveil finite-measurement-induced broadening effects 
in the discontinuous transitions, 
at resonances of the mean quantum {recurrence} time.
{This resonance broadening, encapsulated by the new uncertainty-like relation,
indicates the emergence of dark states in the context of monitored quantum dynamics,
in comparison with the transfer of energy quanta typically observed in spectroscopy
\cite{John2022Spectroscopy_QC_Exp}.
}

Our primary focus is on the $w\to w-1$ transition of the mean quantum return time. 
The $w\to w-1$ transition takes place whenever there are phase factors matching,
hence exhibiting zero width at certain fine-tuning points 
(in theory, under infinite measurements $T_R \to \infty$).
However, due to that only finite measurements are used in real experiments, the restart strategy is by default employed since the data are collected in a finite time span.
Consequently, the original zero-width transition is broadened in experimental observation, allowing the topological transition to be measured.
More specifically, we have found three types of relations.
The first expresses $\langle n_R \rangle$ and $\expval{n}_\text{Con}$ 
in terms of a slow-decay mode of the system, 
namely it shows the dependence on $z_m$ and $T_R$, see Eqs. (\ref{ntr},\ref{resasymp}). 
The second establishes connections between $\langle n_R \rangle$ and $\expval{n}_\text{Con}$
and the variance of the number of attempts till the first detection,
see Eqs. (\ref{nvar},\ref{nrvar}). 
The third relation, as described in Eqs. (\ref{condmean},\ref{rmen}), 
indicates the interplay between the measured mean hitting time and the energies of the system.

{In Ref. \cite{yin2024restart}, 
we verified our theory by tuning the sampling time $\tau$, 
a parameter that is readily implementable on quantum computers. 
However, the system energies involved in our restart uncertainty relations 
are not assumed to be fixed. 
To this end, we consider a chiral quantum walk model 
in which an applied magnetic field breaks time-reversal symmetry, 
thereby introducing conjugate phases acquired by the quantum walker 
when moving forward and backward. 
These phases can be tuned by adjusting the magnetic field, 
which in turn modulates the system energies. 
Consequently, 
we employ this model to investigate resonance broadening 
as a function of the magnetic flux, 
whose variations can induce either degeneracies or splitting of energy levels.
This further validates our theory, 
demonstrating its universal applicability to systems influenced by external perturbations.
}

A natural inquiry is to examine the impact of system size on our primary findings. 
To address this, we analyzed several graph models.
Generally speaking, 
the system size dependence of broadening is non-universal 
due to the variability in chosen energy levels and differences between long-range models 
like the complete graphs and short-range systems like the rings. 
Specifically,  
when focusing on the merging of two phases, 
corresponding to the highest and ground state energies, 
we observe that $w = 1+ L/2$ ($w=(1+L)/2$) for even (odd) ring models, 
while $w = 2$ for the complete graph.
Assuming $\gamma$ is independent of $L$, 
we find for the min-max condition and the complete graph that, the width of the resonances decreases as $L$ increases.
Interestingly, when considering the resonance related to 
the first excited state and the ground state for the ring model, 
we discover that the resonance width actually increases as the system size grows. 
For the complete graph, which has only two energy levels, 
this choice of energy levels is equivalent to the aforementioned min-max scenario.
The key factor determining the broadening effect is how the energy spacing
and the parameter $\lambda$ scale with the system size, as demonstrated in the segment and bipartite graph models. 
The value of $w$ is influenced by the system's symmetry and the degeneracy of energy levels. 
Notably, our time-energy-like restart uncertainty relation 
successfully captures all these diverse behaviors, 
demonstrating its robustness and versatility across different system configurations.
In future studies, the topological transitions from $w$ to $w-2$ or $w-3$, etc., are worth investigating \cite{yin2019}. 
These transitions, as more eigenvalues of the survival operator $\cal{S}$ (as in the Fig.~\ref{fig:Tridip}(c))
approach the unit circle, hold the potential to introduce significant effects into the system.
The study of the broadening effects in the presence of noise could also be of interest, 
for example when the measurement protocol is non-stroboscopic \cite{David2021,Ziegler2021,das2022quantum,Majumdar2023a}.
Importantly, the system dynamics becomes involved in the resonance broadening via the overlap $p_k$ in the parameter $\lambda$, 
in contrast to the infinite measurement case where merely the number of energy phases $e^{-i\tau E_k}$ matters.
Thus, the resonance broadening in this work might be employed to probe system dynamical transition, 
say the localization-delocalization transition of system eigenstates induced by disorders \cite{Yajing2023,EversRMP2008}.

\begin{acknowledgments}
We acknowledge the support of Israel Science Foundation’s grant 1614/21.
\end{acknowledgments}

\section*{CONFLICT OF INTEREST}
The authors have no conflicts to disclose.

\section*{DATA AVAILABILITY}
The data that support the findings of this study are available from the corresponding authors upon reasonable request.


%

\appendix

\section{Rigorous proof of restart uncertainty relation}\label{sec6}

We will provide rigorous proof of the above uncertainty relations.
To do so we will find $F_n$ in the large $n$ limit. We also find an exact expression for $F_n$.
The readers can skip this technical part if not interested in the proof.

\subsection{Explicit formula for $F_n$}
In the following derivation, we note that Eqs.~(\ref{renew}-\ref{neqw}) are not new.
The expression inside the bracket in Eq. (\ref{Fn}) can be rewritten as \cite{Friedman2017a},
%
\begin{equation}
    \phi_n = \bra{\psi_\text{d}}\hat{U}(n \tau)\ket{\psi_\text{d}}
    -
    \sum_{m=1}^{n-1}\bra{\psi_\text{d}}\hat{U}((n-m) \tau)\ket{\psi_\text{d}} \phi_m.
    \label{renew}
\end{equation}
Here $\phi_n$ is the first detection amplitude,
and $F_n=\abs{\phi_n}^2$.
Eq. (\ref{renew}) is also called the quantum renewal equation.
$\ket{\psi_\text{d}}$ is the initial and also the target state of the quantum walker, 
as mentioned before in the Model (Sec. \ref{sec2}). 
In our examples, the target state is a node on the graph, 
and since we have in these examples translational invariance, any node will hold.
%
%
Since Eq.~(\ref{renew}) has a convolution term,
applying the $Z$ transform, namely,
\begin{equation}
    \Tilde{\phi}(z) := \sum_{z=1}^{\infty}z^n \phi_n,
    \label{ztrans}
\end{equation}
we obtain the generating function \cite{Friedman2017a}
\begin{equation}
    \Tilde{\phi}(z) = \frac{\bra{\psi_\text{d}}\hat{\mathcal{U}}(z)\ket{\psi_\text{d}}}{1 + \bra{\psi_\text{d}}\hat{\mathcal{U}}(z)\ket{\psi_\text{d}}},
    \label{generfunc}
\end{equation}
where $\hat{\mathcal{U}}(z) := \sum_{n=1}^{\infty}z^n \hat{U}(n \tau) 
= z e ^{-iH \tau}/(1-ze^{-iH\tau}) $.
The generating function is a useful tool with which we may obtain many results, 
the inversion formula 
 \begin{equation}
     \phi_n = \frac{1}{2 \pi i}\oint_{|z|=1} \frac{dz}{z^{n+1}}\Tilde{\phi}(z)
     \label{inversion}
 \end{equation}
provides a formal solution to the problem. 
Via spectral decomposition of Eq. (\ref{generfunc}) (into the energy eigenbasis), 
we have \cite{Friedman2017a}
\begin{equation}\label{spec}
    \tilde{\phi}(z) = 
    {\sum_{k=1}^w \sum_{l=1}^{g_k} |\bra{\psi_\mathrm{d}}\ket{E_{kl}}|^2
    {ze^{-iE_k\tau}/(1-ze^{-iE_k\tau})} 
    \over 
    \sum_{k=1}^w \sum_{l=1}^{g_k} |\bra{\psi_\mathrm{d}}\ket{E_{kl}}|^2 (1-ze^{-iE_k\tau})^{-1}},
\end{equation} 
where $w$ is 
the number of distinct energy phase factors $\exp(-iE_k\tau)$ 
with non-zero overlap $\sum_{l=1}^{g_k} |\bra{\psi_\mathrm{d}}\ket{E_{kl}}|^2$, 
$g_{k}$ is the degeneracy of $E_{k}$ ($g_k\ge2$ means degenerate energy levels),
and $\ket{E_{kl}}$ are the eigenstates corresponding to $E_k$.
Eq. (\ref{spec}) can be rewritten as
\begin{equation}\label{phiND}
\begin{aligned}
    \tilde{\phi}(z) &= {{\cal N}(z) \over {\cal D}(z)}, \\
    \text{with } 
    {\cal N} (z) 
    &=  z \sum_{k=1}^w \sum_l^{g_k} \left| \bra{\psi_\mathrm{d}}\ket{E_{kl}} \right|^2
        \prod^w_{j=1,j\neq k} \left( z - e^{iE_j\tau} \right), \\
    {\cal D} (z)
    &=  \sum_{k=1}^w e^{iE_k\tau} \sum_l^{g_k} 
        \left| \bra{\psi_\mathrm{d}}\ket{E_{kl}} \right|^2
        \prod^w_{j=1,j\neq k} \left( z - e^{iE_j\tau} \right).
\end{aligned}
\end{equation}
%
%
And one can prove the relation \cite{Friedman2017a,yin2019}
%
\begin{equation}
    {\cal D}(z) = (-1)^{w-1} e^{i\chi} z^w \left[ {\cal N} \left( 1/z^* \right) \right]^*,
\end{equation}
where $\chi = \sum_{k=1}^w \tau E_k$, and the superscript "$\ast$" means conjugate.
Then we can factorize $\tilde{\phi}(z)$ as \cite{Gruenbaum2013,Friedman2017a}
%
\begin{equation}
    \tilde{\phi}(z) = ze^{-i \chi} \prod_{i=1}^{w-1}\frac{z-z_i}{z^*_i(z-1/z^*_i)} ,
    \label{factor}
\end{equation}
where
$\{z_i\}$ are the zeros of ${\cal N}(z)$ or $\tilde{\phi}(z)$.
These zeros are used in Eqs. (\ref{varzeros},\ref{zmmax}) above,
and are located inside the unit circle in the complex plane.
As mentioned, they are also the conjugate of the eigenvalues 
of the survival operator ${\cal S}=\left( 1-\hat{D}\right) \hat{U}(\tau)$. 
This can be proven by applying the matrix determinant lemma 
to the characteristic polynomial of ${\cal S}$, namely,
\begin{equation}
\begin{aligned}
    0=&\text{det} \left[ \eta \mathbf{1} - {\cal S} \right]
    = \text{det} \left[ \eta \mathbf{1} - \hat{U}(\tau) 
    + \ket{\psi_\text{d}}\bra{\psi_\text{d}} \hat{U}(\tau) \right] \\
    =& {\eta} \,\text{det} \left[ \eta \mathbf{1} - \hat{U}(\tau) \right]
     \bra{\psi_\text{d}} \left[ \eta \mathbf{1} - \hat{U}(\tau) \right]^{-1} \ket{\psi_\text{d}}.
\end{aligned}
\end{equation}
The term $\bra{\psi_\text{d}} \left[ \eta \mathbf{1} - \hat{U}(\tau) \right]^{-1} \ket{\psi_\text{d}}$
can be spectrally decomposed as 
$\sum_{k=1}^w \sum_{l=1}^{g_k} |\bra{\psi_\mathrm{d}}\ket{E_{kl}}|^2 \left[ 1/(\eta-e^{-iE_k\tau}) \right]$,
which, equal to $0$, gives the eigenvalues of ${\cal S}$, inside the unit circle, 
that are conjugate of the zeros of $\tilde{\phi}(z)$ (excluding the trivial zero $z=0$).

We note here that the mean hitting time $\expval{n}$ 
(for infinite measurements, i.e. $T_R = \infty$) can be computed by 
\begin{equation}\label{neqw}
    \expval{n} = {1\over 2\pi i} \oint_{|z|=1} 
    \partial_z \ln \left[ \tilde{\phi}(z) \right]\, {\rm d}z,
\end{equation}
which directly gives $\expval{n}=w$ using Eq. (\ref{factor}).
Namely, the mean $\expval{n}$ is identical to the number of zeros of $\tilde{\phi}(z)$, 
{\em inside the unit disk}. 


Substituting Eq.~(\ref{factor}) into Eq.~(\ref{inversion}) and using the residue theorem yield 
\begin{equation}\label{phi45}
    \phi_n = e^{-i \gamma} \sum^{w-1}_{j=1} \left( z_j^{*} \right)^{n-1}
            \left( \frac{1}{z^*_j}-z_j \right) 
            \prod_{k\neq j} \frac{z_j^*(1/z^*_j-z_k)}{z^*_k-z^*_j}
            .
\end{equation}
%
Let $z_j=\rho_j \exp\left( i \theta_j \right)$, 
i.e. $\rho_j = |z_j|$,
$\theta_j = \text{arg}(z_j)\in [0, 2\pi)$, 
and further simplification gives 
\begin{equation}
    F_n = \left| \phi_n \right|^2 
    = \sum_{j,k=1}^{w-1} 
    \frac{\alpha_j \alpha^*_k}{\beta_j \beta^*_k} \,
    \left( \rho_j \rho_k \right)^{n} e^{in \Theta_{jk}},
    \label{Fnab}
\end{equation}
where
$\Theta_{jk} = \theta_k-\theta_j \in [0, 2\pi)$, and 
\begin{equation}
    \frac{\alpha_j}{\beta_j} 
    = \frac{\prod_{i} \left( 1/z^*_i \right)
    \left( 1/z^*_j-z_i \right)}{\prod_{i\neq j} \left( 1/z^*_j-1/z^*_i \right)}
    .
\end{equation}
Hence Eq.~(\ref{Fnab}) has $(w-1)^2$ terms. 
Due to the invariance under the switching between $j$ and $k$ in Eq. (\ref{Fnab}),
the fact that $F_n$ is real is guaranteed by the appearance of paired conjugate terms.
Combining these conjugate off-diagonal terms of Eq.~(\ref{Fnab}) gives
\begin{equation}
    F_n = \sum_{k=1,l>k}^{w-1} 
    \left[
    a_k\rho_k^{2n}+2a_{kl}(\rho_k\rho_l)^{n}\cos (n \Phi_{kl}) 
    \right],
    \label{Fncos}
\end{equation}
where 
\begin{equation}\label{fnparameter}
\begin{split}
    a_{kl} = \left|\frac{\alpha_k \alpha^*_l}{\beta_k \beta^*_l} \right|,\; 
    a_k = \left|\frac{\alpha_k}{\beta_k} \right|^2, \;
    \Phi_{kl} = \Theta_{kl} + \frac{1}{n} 
    \arg \left(\frac{\alpha_k \alpha^*_l}{\beta_k \beta^*_l}\right).
\end{split}
\end{equation}
Hence Eqs. (\ref{Fncos},\ref{fnparameter}) provide a final explicit formula to calculate $F_n$.

\subsection{Behaviors of $F_n$'s tail close to (pseudo)degeneracies: proof for Eq. (\ref{defFn})}
%
We assume that there exists a unique maximum of the set $\{|z_i|\}$ in Eq. (\ref{zmmax}). 
It is in principle possible,
that one may find a couple or more $z_i$'s all being the largest
in absolute value sense \cite{yin2019}. 
We will treat these special cases in future work.
Specifically, we can further simplify Eq. (\ref{Fncos}) to obtain the tail of $F_n$
\begin{equation}\label{eq43}
    F_n \sim a_m \rho_m^{2n} = a_m \left| z_m \right|^{2n}.
\end{equation}
Using Eq. (\ref{fnparameter}), and $\rho_m=|z_m| = 1-\varepsilon \to 1$,
we have
\begin{widetext}
\begin{equation}\label{eq44}
    a_m = \left| {\alpha_m \over \beta_m} \right|^2 
    = \left| {1 \over z_m^*} \left( {1 \over z_m^*} - z_m \right) \right|^2
    \sideset{}{'}\prod_i 
    \left| { 1/z^*_m-z_i \over \left( z^*_i \right) \left( 1/z^*_m-1/z^*_i \right)} \right|^2
    \sim \left( 1 - |z_m|^{-2} \right)^2 
    \sideset{}{'}\prod_i
    \left| { 1-z_i \over z_i^* -1} \right|^2
    \sim \left( 1 - \rho_m^{2} \right)^2
    ,
\end{equation}
\end{widetext}
where $\prod^\prime_i$ means multiplication over all $i$ except for $i=m$.
Therefore, we get a universal formula for $F_n$'s tail, 
in the vicinity of the transition or phase factors matching, namely,
\begin{equation}\label{eq53}
    F_n \sim \left( 1-\rho_m^{2} \right)^2 \rho_m^{2n} 
    = \left( 1 - |z_m|^{2} \right)^2 |z_m|^{2n},
\end{equation}
which confirms rigorously the validity of Eqs. (\ref{defFn},\ref{azm}).
We have assumed that a gap exists between the maximum $|z_m|$ and other zeros of ${\cal N}(z)$ in the system. This holds true for finite systems.
Finally, with Eq. (\ref{eq53}) we derive our main results in Eqs. (\ref{ntr},\ref{resasymp}).
We want to note again that $z_m$ is unique.

\end{document}